\documentclass[fleqn,10pt]{wlscirep}
\usepackage{graphicx}  
\usepackage{dcolumn}   
\usepackage{bm}        
\usepackage{amssymb}   
\usepackage{soul}
\usepackage[normalem]{ulem}

\usepackage{nicefrac}
\usepackage{tikz}
\usetikzlibrary{arrows}
\usepackage{stackengine}

\usepackage{caption}
\usepackage{subcaption}
\usepackage[percent]{overpic}
\usepackage[normalem]{ulem}
\usepackage{amsmath}

\newcommand{\YM}[1]{{\color{black}#1}}

\newcommand{\onlinecite}[1]{\hspace{-1 ex} \nocite{#1}\citenum{#1}}
\begin{document}



\title{\YM{How to Measure} the Entropy of a Mesoscopic System via Thermoelectric Transport}
\date{\today}

\author[1,2,a]{Yaakov Kleeorin}

\author[3,4]{Holger Thierschmann}

\author[4]{Hartmut Buhmann}

\author[5,6,7,8]{Antoine Georges}

\author[4]{Laurens W. Molenkamp}

\author[1,9,b]{Yigal Meir}
\affil[1]{Department of Physics, Ben-Gurion University of the	Negev, Beer Sheva 84105, Israel}
\affil[2]{Center for the Physics of Evolving Systems, Biochemistry and Molecular
	Biology, University of Chicago, Chicago, 60637, USA}
\affil[3]{Kavli Institute of Nanoscience, Faculty of Applied Sciences, Delft University of Technology, Lorentzweg 1, 2628 CJ, Delft, The Netherlands}
\affil[4]{Physikalisches Institut (EP III), Universität Würzburg
	D-97074 Würzburg, Germany}
\affil[5]{Centre de Physique Theorique, Ecole Polytechnique, CNRS,Universite Paris-Saclay, 91128 Palaiseau, France}
\affil[6]{College de France, 11 place Marcelin Berthelot, 75005 Paris, France}
\affil[7]{Center for Computational Quantum Physics, Flatiron Institute,
	162 Fifth Avenue, New York, New York 10010, USA}
\affil[8]{DQMP, Universite de Geneve, 24 quai Ernest Ansermet, CH-1211 Geneve, Switzerland}
\affil[9]{The Ilse Katz Institute for Nanoscale Science and Technology, Ben-Gurion University of the
	Negev, Beer Sheva 84105, Israel}

\affil[a]{Kleeorin@gmail.ac.il}

\affil[b]{Ymeir@bgu.ac.il}

\begin{abstract}
Entropy is a fundamental thermodynamic quantity indicative of the accessible degrees of freedom in a system.  While it has been suggested that  the entropy of a mesoscopic system can yield nontrivial information on emergence of exotic states, its measurement in such small electron-number system is a daunting task. Here we propose a method to extract  the entropy of a \YM{Coulomb-blockaded} mesoscopic system from  transport measurements. We prove analytically and demonstrate numerically the applicability of the method to \YM{such} a mesoscopic system of arbitrary spectrum and degeneracies. We then apply our procedure to measurements of thermoelectric response of a single
quantum dot, and demonstrate how it can be used to  deduce the entropy change across Coulomb-blockade valleys, resolving, along the way, a long standing puzzle of the experimentally observed finite thermoelectric response at the apparent particle-hole symmetric point.
\end{abstract}

\flushbottom
\maketitle
\section*{Introduction}
The entropy of a mesoscopic system can yield nontrivial information on emergence of exotic states, such as two-channel Kondo impurity \cite{andrei1984}, non-abelian anyons in the $\nu=5/2$ regime \cite{ben-shach2013,Viola2012} or Majorana modes in topological superconductors \cite{smirnov2015}. Nevertheless, the measurement of entropy in such small electron number systems is highly nontrivial. \YM{Refs.~\onlinecite{Cockins2010} and \onlinecite{Hofmann2016} used the asymmetry of the in and out tunneling processes in a quantum dot (QD) to deternmine the degeneracy of the QD states},  while recent elegant experiments \cite{Hartman2018} have  employed the thermodynamic Maxwell relation between entropy evolution and chemical potential,  $(\partial\mu/\partial T)_{n}=-(\partial S/\partial n)_{T}$, in order to directly measure entropy transitions in semiconductor QDs. \YM{This latter experiment} required  measurements of another thermodynamic quantity  - the charge of the system as a function of gate voltage, for different temperatures, and hence a specially designed device. Here we propose a different approach to this problem: can one extract information about the entropy from  transport measurements ? Obviously, this requires a measurement of both particle and thermal (entropy/heat) transport.
This question has been addressed in the context of bulk solids \cite{Chaikin1976,Behnia2004,Zlatic2007,Mravlje2016}, with sometimes debated points of view. A general relation exists between the low-temperature thermopower and
specific-heat (entropy) of a free electron gas, and this relation appears to apply in a number of
materials\cite{Behnia2004,Zlatic2007}. However, thermopower is, quite generally, a transport coefficient
and its relation to entropy has been shown to be questionable in systems with strongly anisotropic transport
for instance\cite{Mravlje2016}. In the opposite high-temperature limit, where temperature is the largest energy scale in the system, general relations between the thermopower and derivatives of the entropy can be derived, embodied in the Heikes \cite{R.R.HeikesandR.W.Ure1961,Chaikin1976,Doumerc1994} and Kelvin \cite{Peterson2010,Mravlje2016} formulas.



\YM{Consider an arbitrary mesoscopic system in the Coulomb-blockade regime (where only $N$ and $N+1$-particle states are energetically relevant), whose entropy one wishes to measure.
The method we propose here is based on a general observation, which is also an important result of our work: if one weakly couples this system to leads, } the conductance  of such an interacting system can be put in the form of a non-interacting conductance formula, provided one takes into account a temperature-dependent shift of the chemical potential (gate voltage). \YM{The thermal response  (TR), in turn, can be written in a similar manner, where the temperature-dependent shift in the chemical potential produces an extra contribution. We show that this extra term}, which can be determined by comparing the actual thermal response of the system to that of the related non-interacting system (which can be estimated using a newly introduced high-temperature version of the \YM{original} Mott formula\cite{Cutler1969}),  can be used to extract the entropy \YM{of such a mesoscopic system} even in the case of arbitrary spectrum and degeneracies, and then  demonstrate the usefulness of the approach by applying it to several model systems.
One big advantage of our formulation is that one can apply it to any \YM{such} mesosopic system where measurements of both electrical conductance and thermopower are available. This allows us to apply our procedure to existing data of thermoelectric response of a single QD, and demonstrate how it can be used to  deduce the entropy change and the QD's degeneracy. In the process we explain the long standing puzzle of the observation of a non-zero thermopower at the apparent electron-hole symmetry point in the Coulomb Blockade (CB) valley \cite{Scheibner2005,Svilans2018}.
\section*{Results}
\subsection*{General Formulation}
Consider a general mesoscopic system with many-body eigenstates $\Psi^{(N)}_i$, where $N$ is the number of electrons in that state, with energies $E^{(N)}_i$ \YM{(with $g^{(N)}_i$  the degeneracy of the energy $E^{(N)}_i$), whose entropy one wishes to measure. In order to perturb the system as little as possible, we weakly couple the mesoscopic system to two reservoirs \YM{(with  coupling $V_i$ for each state $i$)}. In this weak-coupling limit  $\Gamma_{ij}=2\pi V_i V_j \rho$,   the characteristic level broadening, with $\rho$ the density of states in the reservoirs, obeys $\Gamma_{ij}\ll T$, where $T$ is the temperature. In this limit} the conductance $G$ through the mesoscopic system can be written as the sum of individual transitions from state $i$ with $N$ electrons to state $j$ with $N+1$ electrons \cite{Meir1992a}
\begin{equation}
G(\mu,T)=\sum_{ij} G_{ij}(\mu,T)=\sum_{ij}\mathcal{T}^{(0)}_{ij}\times \left[(P_i^{(N+1)}(\mu,T)+P_j^{(N)}(\mu,T)\right]\frac{d f(E^{(N+1)}_i-E^{(N)}_j-\mu,T)}{d\mu}
\label{eq:G}
\end{equation}
where $\mathcal{T}^{(0)}_{ij}$ is equal to  $\Gamma_{ij}$ times the overlap of the $N+1$-particle many-body wave function   $\Psi^{(N+1)}_j$ with the $N-$particle wavefunction $\Psi^{(N)}_i$ , with the addition of the electron tunneling in from the leads (or the reverse process) (see Supplementary Information, Eq.~1). In the above $f(E,T)$ is the equilibrium Fermi function, $\mu$ the chemical potential, and $P_i^{(N)}(\mu,T)=e^{-(E_i^{(N)}-\mu N)/T}/Z$  is the equilibrium probability of the system to be in the $N$-particle many-body state $i$, with $Z$ the partition function \YM{(except for the experimental part, we use $k_B=1$ throughout the paper, where $k_B$ is the Boltzman coefficient, so that temperature has units of energy and entropy is dimensionless)}.  A similar expression can be written for the TR, defined as $dI/dT$, the change in the linear-response current due to temperature difference between the leads, in analogy to conductance, with $df/d\mu$ being replaced by $df/dT$. \YM{We assume that the Coulomb energy is significantly larger than $T$ and $\Gamma$ so that for a given chemical potential, $G$ involves transitions between states with only $N$ or $N+1$ particles.}
A crucial step in our formulation is the demonstration that the above general expressions for the conductance and the thermal response for an arbitrary interacting system  can be accurately written, in the vicinity of each $N \rightarrow N+1$ transition, as those for a non-interacting system, but with a temperature-dependent effective chemical potential (see Supplementary Note 1):
\begin{equation}
G_{ij}(\mu,T)=C(T)G^{NI}_{ij}(\mu+\Delta_{ij}(T),T)
\label{eq:GijNI}
\end{equation}
where $G^{NI}_{ij}$ is the conductance for a non-interacting system with same spectrum and couplings, and $C(T)$ is some temperature dependent prefactor, that will drop out when the relation between G and TR is derived. This temperature dependent shift in the chemical potential is  given by
\begin{equation}
\Delta_{ij}(T)=\frac{E^{(N+1)}_{j}-E^{(N)}_{i}}{2}+\frac{T}{2}\log\big[\frac{\sum_j g^{(N+1)}_j e^{-E^{(N+1)}_{j}/T}}{\sum_i g^{(N)}_{i} e^{-E^{(N)}_{i}/T}}\big]
\end{equation}
In the simple case of a transition from an empty state into  a single  level, with degeneracy $g$, this shift reduces to $\frac{1}{2}T\log g$, which has been noticed before \cite{Beenakker1991,Viola2012}, and has been measured experimentally \cite{Cockins2010}. In that case this shift was attributed to the fact the chemical potential has to shift in order to compensate for the fact there are $g$ ways for an electron to tunnel into the QD, while having a single channel for tunneling out, an asymmetry that has been verified experimentally \cite{Cockins2010,Beckel2014}. In contrast, our expression indicates that in the case of many levels, which has not been discussed before, the temperature-dependent part of the shift does not depend on which level the electron tunnels through, and what its degeneracy is. This part of the shift is identical for all transitions, and is equal  one half of the difference of the canonical free energies between the CB valleys corresponding to $N$ and $N+1$ electrons.

The  explicit dependence of $\Delta_{ij}$ on $T$ allows us to write, in a similar manner to Eq.~\ref{eq:GijNI},   an explicit expression for the TR of a general interacting system in terms of its conductance  and the TR of the related non-interacting system,
\begin{equation}	
\mathrm{TR}_{ij}(\mu,T)=C(T)\mathrm{TR}_{ij}^{NI}(\mu+\Delta_{ij}(T),T)+G_{ij}(\mu,T)\Delta_{ij}(T)/T
\label{eq:general}
\end{equation}
\YM{In order to derive an equation for $\mathrm{TR}^{NI}$, the thermal response of a non-interacting system with same spectrum and couplings, we generalize the Mott formula \cite{Cutler1969}, valid for $T\ll\Gamma$, to the regime $T\gg\Gamma$ (see Eq.~\ref{HTM} in the Methods section and Supplementary Note 2 for derivation).  Thus, the deviation of the $\mathrm{TR}$ from $\mathrm{TR}^{NI}$ (calculated from the conductance) allows us to estimate $\Delta_{i,j}(T)$,  and  consequently the entropy difference between the consecutive CB  valleys:
$\Delta S_{N \rightarrow N+1}=2d\Delta_{ij}(T)/dT$.}

 So, given the experimentally or numerically obtained $G(\mu,T)$ and $TR(\mu,T)$, the procedure we propose for finding the entropy difference between consecutive CB valleys is the following: 1. Given $G(\mu,T)$, one can use our variant of the Mott formula (Eq.~\ref{HTM} in the Methods section) to evaluate the first term on the right-hand side (RHS) of  Eq.~\ref{eq:general}. 2. For a given temperature, the difference between this term and the  actual TR, which is a function of the chemical potential, is proportional to $G(\mu,T)$. We denote this proportionality constant A(T) (note that $A(T)$ is the only fitting number required, for a given temperature, to map the two functions on top of each other).
3. Given the obtained $A(T)$, the difference in entropy between the valleys is then given by $\Delta S_{N \rightarrow N+1}=2d\left[T\times A(T)\right]/dT$. (A step by step description of the fitting process is detailed in Supplementary Note 3).

 In the following we demonstrate the usefulness of this formalism in model systems, where one can compare the entropy obtained using the above relation to that calculated directly from thermodynamic considerations, and finally we apply our formalism to available experimental data.

\subsection*{Comparison to numerical calculations}

\begin{figure}[h]
	\includegraphics[width=1\textwidth]{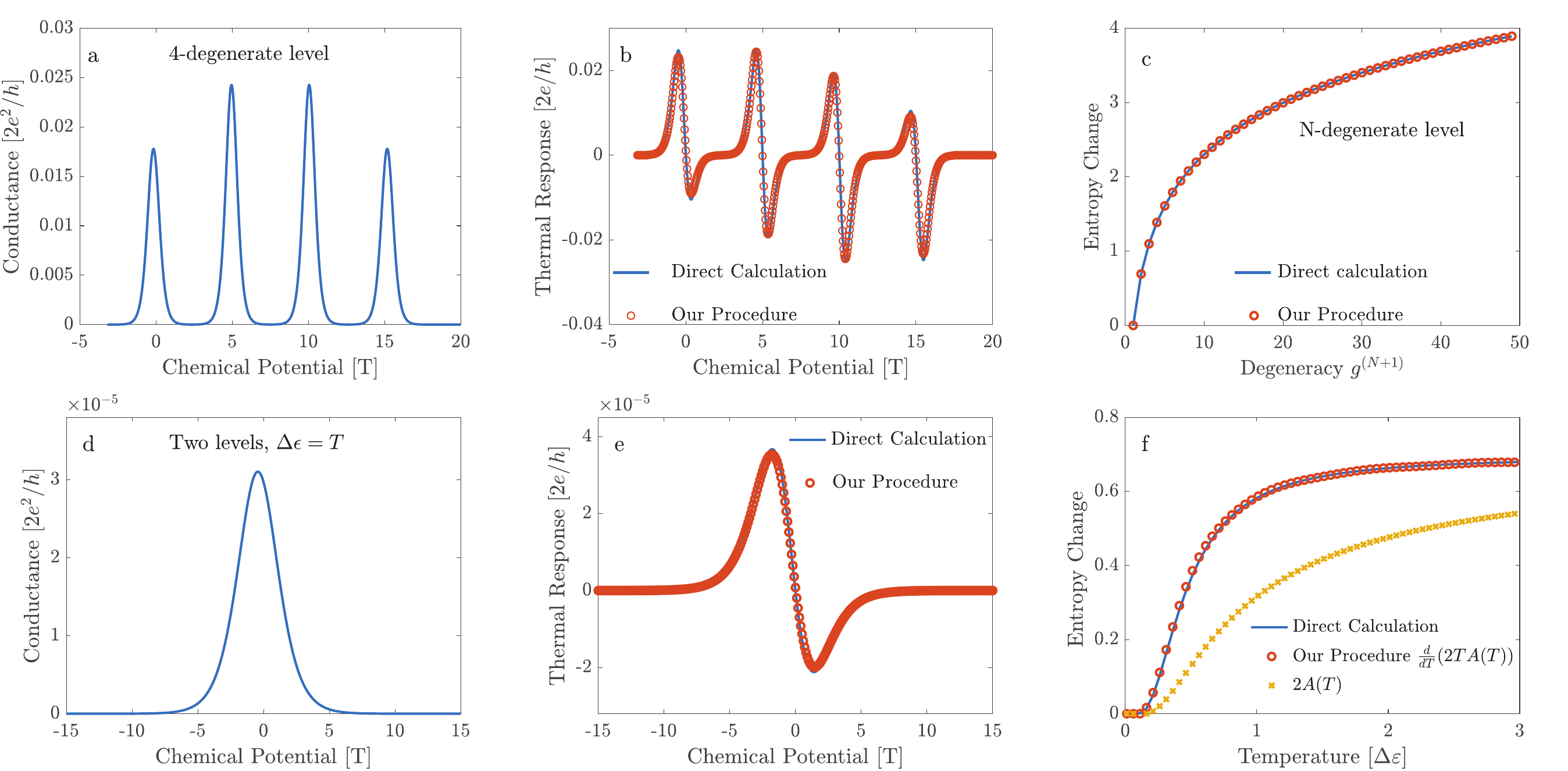}

	\caption{Demonstration of the fitting Procedure. (a,b) Transport coefficients through a  fourfold degenerate quantum dot, calculated via Eq.~\ref{eq:G}: (a) Conductance, (b) TR (solid blue line) with comparison to the derived expression [Eq. \ref{eq:general}] (red circles). The degeneracies for $n=0,1,2,3,4$-electron many-body states are $g^{(N)}=1,4,6,4,1$, respectively \YM{(Each peak was separately fitted).} (c) Entropy change between two valleys with first valley degeneracy $g^{(N)}=1$, as a function of second valley degeneracy $g^{(N+1)}$, calculated using the proposed procedure (red circles) compared to the exact result $\log g^{(N+1)}$(solid blue line).
		(d,e,f) Transport through a $U\rightarrow \infty$ QD with 2 single-particle non-degenerate interacting  levels, separated by $\Delta \epsilon=T$, calculated  via Eq.~\ref{eq:G}: (d) Conductance, (e) TR (solid blue line) with comparison to the derived expression [Eq. \ref{eq:general}] (red circles).
		(f) Entropy change between the two valleys as a function of temperature. Direct thermodynamic calculation of entropy change (solid blue line) is compared to  our procedure ($d2T \times A(T)/dT$) (red circles). $A(T)$ is shown as yellow crosses.}
	
	\label{fig:TRexample}
\end{figure}

Let us start with a simple example where in each \YM{$N$-electron subspace} there  are $g^{(N)}$ degenerate $N$-particle states of energy $E^{(N)}$, and all other states can be ignored (i.e. the level spacing is much higher than temperature). In this case the entropy $S_N$ in each valley is equal to $\log g^{(N)}$, and is temperature independent. In this case, \YM{one indeed finds that the proportionality constant is temperature independent}, $A(T)=\log(g^{(N+1)}/g^{(N)})/2$. Fig.~\ref{fig:TRexample}b illustrates the correspondence between the TR obtained \YM{directly}, using Eq.~\ref{eq:G}, 
and that obtained by the RHS of Eq.~\ref{eq:general} (red circles), for a four-fold degenerate interacting QD, relevant, for example, to a carbon nanotube QD (see also experimental section below).  \YM{The conductance used in evaluating both terms in the RHS of Eq.~\ref{eq:general} was also obtained via Eq.~\ref{eq:G} }(and is shown in Fig.~\ref{fig:TRexample}a). In this case there are 4 CB peaks, \YM{separating valleys} with degeneracies $g^{(N)}=1, 4,6,4$ and $1$ for $N=0, 1, 2, 3$ and $4$. In order to construct the estimate for the TR in Fig.~\ref{fig:TRexample}b we have \YM{used the above fitting procedure separately for each peak, as the entropy difference between consecutive valleys is different for each peak}.
The figure displays an almost perfect agreement between the \YM{direct calculation }
 of the TR and that obtained by our Ansatz.

In this case, as the entropy change $\Delta S$ between the valleys is temperature independent,  the estimate of $A$ at a single temperature is directly proportional \YM{ to the entropy change through $\Delta S=2A$}. In particular, the entropy change across the first CB peak is a direct measure of the degeneracy of the QD ($4$ in the above example). We have repeated the procedure for QDs of arbitrary degeneracy. Fig.~\ref{fig:TRexample}c depicts the entropy change  deduced using our procedure (red circles), compared to the expected change in entropy($\log g^{(N+1)}$). We see a perfect agreement even up to large degeneracies.  As mentioned above, some aspects of this simple case of a single degenerate level have been addressed before, and it has been suggested that the thermopower through a single-level QD can be used, e.g., to deduce the nature of the neutral modes in the fractional quantum Hall regime \cite{Viola2012}.

The advantage of our procedure lies in its application to a multi-level mesoscopic system, such as a multi-level QD, or to a multi-dot system, where the entropy is temperature dependent. As an example, let us  consider the case of two singly degenerate levels, with level spacing $\Delta \epsilon$ (describing, for example, a single-level QD in a magnetic field). One expects that when $T\ll \Delta \epsilon$ the entropy of the single-electron system will be equal to zero, while for higher temperature, larger than $\Delta \epsilon$, it will increase to $\log2$. As the entropy is temperature dependent, one has to perform the procedure for all $T$ in order to extract $A(T)$, its derivative, and consequently the entropy. For simplicity, we assume that the transition through one of the levels dominates the transport, so Eq.~\ref{eq:general}, which corresponds to a transition between specific states, will also reflect the full transport coefficient of the system. As we will demonstrate, even though a single transition dominates the transport, the resulting procedure yields the full entropy change in the system.

Fig.~\ref{fig:TRexample}d and e depict, respectively, the calculated conductance and TR, again  using Eq.~\ref{eq:G}, for a specific temperature, $T=\Delta \epsilon$. Fig.~\ref{fig:TRexample}e also shows the TR derived from our procedure - the fitting leads to  $A(T=\Delta \epsilon )$ for this temperature. Repeating the same procedure for many temperatures, one is able to produce the whole curve $A(T)$, and then the entropy change, $\Delta S =2d\left[T A(T)\right]/dT$. The resulting estimate for the entropy change is plotted in Fig.~\ref{fig:TRexample}f along with the thermodynamic calculation of the entropy change:
\YM{$\Delta S_{N\rightarrow N+1}=-\partial \left[F_{N+1}(T,\mu)- F_N(T,\mu)\right]/\partial T$ with $F_N(T,\mu)$ the free energy of the $N$-electron system}.
Again we observe excellent agreement between the entropy deduced in our procedure and the direct calculation. In Supplementary Note 4 we discuss our procedure for the case when several transitions are relevant to the total transport.

\begin{figure}[h]
	\includegraphics[width=1\textwidth]{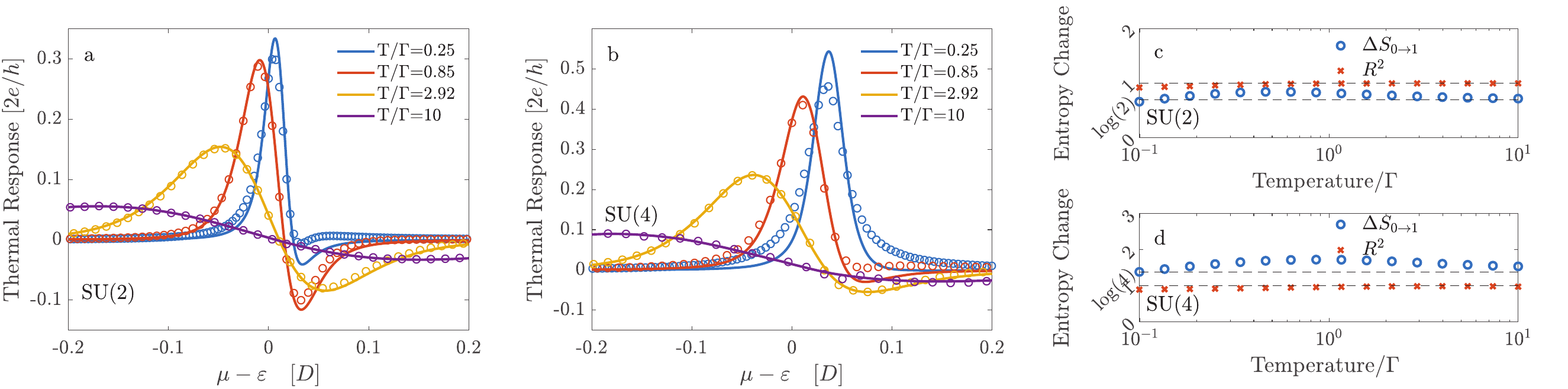}

	\caption{Extension of the procedure to low temperatures. \YM{Fitting of the TR obtained directly from NRG (solid line) with TR obtained from Eq.~\ref{eq:general} (circles), for the (a) two-fold, and (b) four-fold degenerate quantum dot, in the vicinity of the first CB peak, for various temperatures.  (c,d) Calculation of the entropy change  across the first CB peak for a wide range of temperatures for a (c) two-fold, and (d) four-fold degenerate quantum dot, where the expected entropy changes are $\log2$ and $\log4$, respectively.  The closeness of the $R^2$ estimate of the fitting procedure (crosses) to unity indicates the excellent agreement between the two curves of TR, as shown in (a,b). The  $x$-axis in (a,b) is in units of $D$, half the band width in the leads, and  $\Gamma=0.01D$ and $U=D$ in all three panels.}}
	\label{fig:SchemeLowT}
\end{figure}

Interestingly, while this formalism was derived for the \YM{weak coupling ($\Gamma\ll T$)} regime, empirically its validity extends outside this strict regime. Since Eq.~\ref{eq:G} does not apply to the regime $\Gamma\gtrsim T$, we have employed here the numerical-renormalization-group (NRG) method (see Methods), which is accurate down to zero temperature. 
Fig.~\ref{fig:SchemeLowT} demonstrates the validity of our formalism and shows that the estimates of the entropy, using our procedure for the cases of a two-fold ($SU(2)$) and  four-fold ($SU(4)$) degenerate single-level QD, agree with expected values ($\log2$ and $\log4$, respectively),
down to $T\simeq 0.1 \Gamma$.
The fitting procedure that corresponds to Eq.~\ref{eq:general} remains accurate throughout the presented region of temperatures with coefficient of determination ($R^2$) values of close to unity (crosses in Fig.~\ref{fig:SchemeLowT}c,d). \YM{Thus, at least for these two models, our approach extends to couplings to the leads $\Gamma$ which are of the order or even larger than temperature.}

\subsection*{Application to Experiments}

\begin{figure}[h]
	\includegraphics[width=1\textwidth]{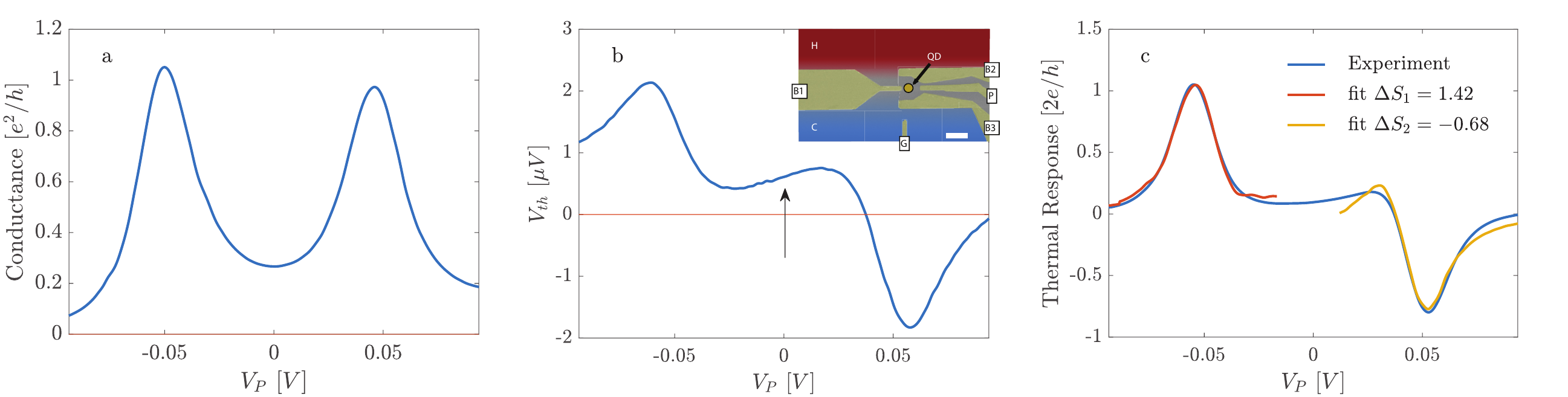}

	\caption{Fitting of the  experimental data. (a,b) Experimental measurements of (a) conductance and (b) thermovoltage through the QD device, depicted in false color in the inset to (b). The horizontal axis corresponds to the QD energy, obtained from multiplying the plunger gate voltage $V_P$ with gate lever arm $\alpha$ (see methods), and shifting the  point of zero energy to the center of the Coulomb blockade valley. The thermovoltage has a  non-zero value in the middle of the valleys around the apparent particle-hole symmetry point (arrow). (c) Fitting procedure [Eq. \ref{eq:general}], performed directly on the experimental data and each peak separately.}
	\label{fig:exp}
\end{figure}

One of the main advantages of our approach, compared, e.g. to that of ref. \onlinecite{Hartman2018},
 is that it can be readily applied to any previous transport experiment in a mesoscopic system,  for which conductance and TR data are available. As an example of the usefulness of the suggested procedure, we have analyzed recent thermoelectric measurement results\cite{thierschmann2014heat} through a QD device,
 formed in a two dimensional electron system  of a GaAs/AlGaAs heterostructure using split-gate technology. This technology allows for a high degree of control over system parameters such as QD energy and tunnel coupling $\Gamma$ between the QD and the reservoirs, by adjusting the voltages applied to the split gates.
The sample is shown in the inset to Fig~\ref{fig:exp}b. Gates B1, B2 and B3 are used to form the QD (yellow dot). The tunnel coupling between the QD and the reservoirs H and C can be controlled symmetrically adjusting the gate voltage applied to gate B1. Gate P, the so-called plunger gate, is used to continuously tune the electrochemical potential of the QD, and consequently the number of electrons on the QD. Gate G is not used in these experiments and is kept at ground at all times.

 The sample is cooled down in a dilution refrigerator, with an electron base temperature of $\approx 230$ mK, in the presence of a small perpendicular magnetic field (B = 0.6 T) \cite{VanderWiel2000}. In order to establish a temperature difference $\Delta T$ across the QD,  a small heating current was applied to reservoir H (see Methods section and Supplementary Note 5), thereby mainly enhancing the electron temperature in that reservoir.
 The thermovoltage  $V_{th}$ is then obtained by recording the voltage drop across the QD as a response to the temperature increase in reservoir H under open circuit conditions (see methods section and Supplementary Note 5 for further details), thus $V_{th}= \mathrm{TR}\times \Delta T /G$.

\begin{figure}[h]
	
	\includegraphics[width=1\textwidth]{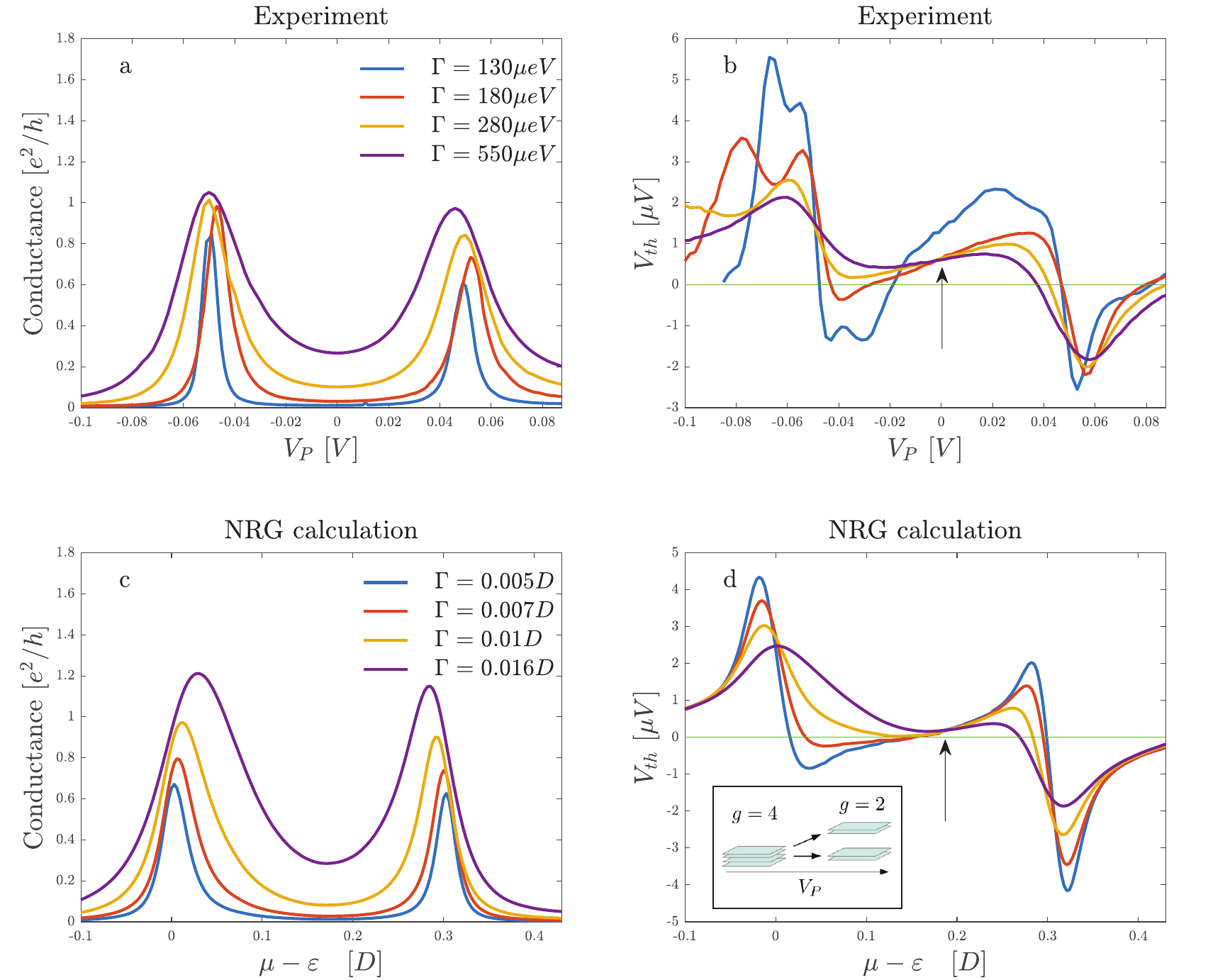}
	\caption{Correspondence between numerical and experimental data for different couplings. Experimental measurements of (a) conductance and (b) thermovoltage through the same device as in Fig.~\ref{fig:exp}, for several values of tunneling widths $\Gamma$. The anomalous nonzero value of the crossing point of the TR curves is denoted by an arrow (due to experimental ambiguity of reference chemical potential, the different curves were aligned so that the apparent particle-hole symmetry point is shifted to $V_P=0$). 
		Theoretical NRG calculations of (c) conductance and (d) thermopower through a QD with two spin-degenerate levels,  with linearly varying level spacing, depicted in the inset to (d). The numerical plots were shifted horizontally so that the minima inside the valley for all plots coincided for alignment as in the experimental plots. The results also indicate a non-zero crossing point (arrow). The  $x$-axes in (c) and (d) as well as $\Gamma$ are in units of $D$, half the band width in the leads, and we used $U=0.3D$.}
	\label{fig:th}
\end{figure}

Fig.~\ref{fig:exp}a and b depict the experimental data for $G$ and $V_{th}$, respectively, for a pair of CB peaks. Interestingly, the data show that at points of apparent particle-hole symmetry in the conductance (e.g. arrow in Fig.~\ref{fig:exp}b and crossing point in Fig.~\ref{fig:th}b),  $V_{th}$ does not vanish as would be expected from the usual, spin-degenerate QD, described by the standard single-impurity Anderson model \cite{Costi2010}. \YM{This experimental observation (see also Refs.~\onlinecite{Scheibner2005,Svilans2018}) is to this day an unresolved puzzle in the field (see Ref.~\onlinecite{Karki2017} for an attempt to resolve this puzzle).}

In the following we detail our analysis of these CB peaks.
\YM{It has been noted before \cite{Svilans2018} that under the condition of heating one reservoir, the actual temperature of the QD can differ greatly from the fridge's temperature. Since in the present case where $T<\Gamma \simeq 550\mu eV$, the actual temperature cannot be deduced from the width of the CB peaks, we use the temperature as an additional fitting parameter. In addition, since the x-axis relation between the conductance measurement (Fig.~\ref{fig:exp}a) and the thermovoltage measurements (Fig.~\ref{fig:exp}b) were not experimentally established, another fitting parameter is introduced: the x-axis shift in the measured conductance relative to the measured thermovoltage.}
The results of  fitting the TR to Eq.~\ref{eq:general} are depicted in Fig.~\ref{fig:exp}c. As can be seen in the figure, there is a good agreement between the fit and the observed TR in the vicinity of each peak, again using only a few fitting parameters to fit the whole curve (see Supplementary Note 3 \YM{for a detailed step-by-step of the analysis of the experimental data using our procedure}), illustrating the experimental validity of our approach. Due to the limited availability of the data we used $G(\mu,T)$  instead of  $G(\mu,\gamma_2 T)$ to estimate $\mathrm{TR}^{NI}$. However, this should make a little difference when $T<\Gamma$.

In applying our method to the experiment, one needs to translate the measured $V_{th}$
to the thermo-electric response TR  by dividing by $\Delta T$. This value, however, is not easily and accurately determined  in an experiment and thus leads to uncertainties in the absolute values of the entropy changes across the peaks. On the other hand, the ratio of these entropy changes across consecutive peaks is  independent of $\Delta T$, and is found to be $-2.07\pm0.13$  for the two peaks depicted in Fig.~\ref{fig:exp} (the errors estimate is due to variation in possible fitting region around the peaks, see Supplementary Note 3). The simplest  scenario  giving rise to such a ratio, is that the entropy change across the first peak is $\log4$ while the second is $-\log2$.
This means that the first peak signals a transition into a four-fold degenerate state, while the second peak may either correspond to a transition from a four-fold degenerate to a two-fold degenerate state, or from a two-fold degenerate state to a non-degenerate state. This suggests a deviation from the naive picture of consecutive filling of a four-fold degenerate state. Including this scenario into our fit, $\Delta T$ is found to be $\approx 20mK$, which is close to the experimental estimate of being of the order of 30 mK (see methods and Supplementary Note 5).

While the degeneracy of these two levels seems fortuitous, such a model, in fact, has been claimed to be generic for transport through QDs \cite{Silvestrov2000,silvestrov2001,Golosov2006}, and has been invoked to explain the repeating phase jumps in the transmission phase through such a dot\cite{yacoby1995, Yacoby1996}. In these works this is caused by two overlapping levels with different tunneling widths. At each conductance valley the narrow level is filled by an additional electron, shifting the energies of the narrow and the wide level differently, thus leading naturally, due to the degeneracy, to the entropy change of $\log4$ across the first peak. In this scenario, after the second conductance peak the narrow level is doubly occupied, and does not play an additional role in transport, while the wide level is shifted up to overlap with another narrow level, and the process repeats itself. This explained the repeated phase change across consecutive conductance peaks\cite{yacoby1995, Yacoby1996}, and is, in fact, consistent with the observation that the upshift of the TR from zero at the apparent particle-hole symmetric point happens in consecutive pairs of conductance peaks \cite{Scheibner2005}.

Experimentally, one can easily change the tunneling rates $\Gamma$ between the QD and the leads through the split gate technique. These data, depicted in Fig.~\ref{fig:th}a and b,  can then be used to differentiate between these possible scenarios.  We found that the model that best reproduces the experimental findings, is that of a QD with two spinful states with an energy difference  $\Delta \epsilon$ that depends on gate voltage \YM{(in the model we used the same $\Gamma$ for both levels to avoid additional parameters)}. Similar evolution of the degeneracy as a function of chemical potential has already been observed in quantum nano-tubes\cite{Pecker2013a}.

In this model, around the gate voltage corresponding to the first peak (QD energy $\epsilon\sim-0.75$meV), the two levels are almost degenerate yielding a net four-fold degeneracy($g_N=0, g_{N+1}=4$) which is lifted as the gate voltage is tuned toward the second peak, around QD energy $\epsilon\sim0.75$meV ($g_N=2, g_{N+1}=1$) (as illustrated in the inset of Fig.~\ref{fig:th}d). This interpretation leads to the observed values of entropy change.

Fig.~\ref{fig:th}c and d depict NRG calculation of a specific model for various values of $\Gamma$, where the energy difference between the levels changes linearly with chemical potential, $\Delta \epsilon=a + b (\mu-\epsilon)$, with $a=-0.01D, b=0.13$ \YM{($D$ the bandwidth of the leads)}. The model reproduces the essential experimental features and those captured by varying $\Gamma$. Some features in the experimental data, such as small side peaks in the lower two values of $\Gamma$, attributed to excited states \cite{Beenakker1992}, are not captured within the current simple model. Interestingly, this model naturally reproduces the non-zero value of the TR at the seemingly particle-hole symmetric point which is also visible in the experimental data (crossing point in Fig.~\ref{fig:th}b, marked by an arrow). This anomalous increase of the TR around the middle of the valley is attributed to a non-trivial degeneracy, thus providing a natural explanation that this value of gate voltage does not correspond, in fact, to a particle-hole symmetric point. (An alternative explanation, based on non-linear effects, was suggested in recent work\cite{Karki2017}.)

\section*{Discussion}
In this work, we have derived a theoretical connection between the entropy and transport coefficients in mesoscopic junctions. This connection relates the TR of a \YM{Coulomb-blockaded} mesoscopic system with arbitrary many-body levels to the conductance and the entropy change between adjacent CB valleys. \YM{While the derivation was introduced for weak coupling $\Gamma$ between the system and the leads (in comparison with temperature), we have demonstrated numerically that, for the case of 2-fold and 4-fold degenerate QD, the method is accurate  also for temperatures well below  $\Gamma$}. This allowed us to apply the method to experimental data in that regime, which yielded non-trivial, and in fact unexpected information about the entropy in each CB valley.
The deduced theoretical model, which described the experimental QD,   reproduced the measured thermopower and resolved the long-standing puzzle of a finite TR in the apparent particle-hole symmetric point.

The success of this procedure suggests possible venues to  extend this analysis \YM{especially towards the study of entropy of exotic states. One direction would be to extend the method to low temperatures, thus enabling the determination the degeneracy of the ground state of the full system. This, for example, is particularly relevant to exotic phases, such as the two-channel Kondo system, where the zero temperature entropy is non zero. If the TR of this system can be utilized to deduce the entropy of the ground state, this can be a smoking gun for the observation of the two channel Kondo ground state \cite{Potok2007} or other such non Fermi liquid ground states. Such an extension has also been suggested in parallel by Sela et al. \cite{Sela} to measure the fractional entropy of Majorana zero modes.
}

\thispagestyle{empty}
\section*{Methods}
\subsection*{High Temperature Mott Relation} In relating the non-interacting conductance and TR we use a high temperature adaptation of the Mott relation\cite{Cutler1969}.\YM{
\begin{equation}	
\mathrm{TR}^{NI}(\mu,T)=\gamma_1 T \frac{dG^{NI}(\mu,\gamma_2 T)}{d\mu},
\label{HTM}
\end{equation}
where the superscript $NI$ denotes a non-interacting system, and $\gamma_2=2/\sqrt{3}, \gamma_1=2 \gamma_2^3$ are universal values related to properties of the Fermi function (for derivation see Supplementary Note 2). }
\\

\subsection*{Numerical Renormalization Group} for the density-matrix numerical renormalization group (DM-NRG) results we used the open-access Budapest Flexible DM-NRG code\cite{Toth2008,Legeza2008}. The  expectation values and the transmission spectral function, required for the evaluation of the conductance through the double dot device \cite{Meir1992a}, were calculated, assuming, for simplicity, equal couplings to the left and right leads, $\Gamma=\pi \rho V^2$, and equal and constant density of states $\rho=1/2D$ in the two leads, with a symmetric band of bandwidth $2D$,  around the Fermi energy.
\YM{The NRG simulation is able to output the many body discreet energy states that the system can occupy and their respective spectral weight, $\epsilon_i, w_i$. Transport coefficient are then calculated using $G(\mu,T)=\Gamma \pi \sum w_i df(\epsilon_i-\mu,T)/d\mu$  and  $\mathrm{TR}(\mu,T)=\Gamma \pi \sum w_i df(\epsilon_i-\mu,T)/dT$.}
\\

\subsection*{Experiment} Our sample is designed similar to the one used by Scheibner et al. \cite{Scheibner2005}. The electron reservoir H which serves as a hot lead for the quantum dot in our thermopower experiments is shaped into a channel of width $w = 2 \mu m$ and length $l = 20 \mu m$ (see Supplementary Fig.~ 5). The QD is situated on one side of the channel, delimited by gates B1 and B2,while the opposite side of the channel is delimited by the two gates Q1 and Q2, forming a quantum point contact (QPC) which is positioned exactly opposite to the quantum dot. The QPC is adjusted to the conductance plateau at G = 10 $e^2/h$. It separates the heating channel H from the reservoir REF which is kept at ground potential. At the two ends of the heating channel (separated by the distance $l = 20 \mu m$) the 2DES opens up quickly into large reservoirs. The channel can be contacted electrically through two Ohmic contacts $I_1$ and $I_2$. We apply a heating current $I_h = 70 nA$ to the channel, which is modulated at a low frequency $\omega = 13$ Hz. Because at low temperature electron-electron scattering is the dominant scattering mechanism on length scales up to several 10 $\mu$m in our system, the power $P_h$ introduced through $I_h$ is dissipated inside the channel only into the electron gas while in the larger reservoirs outside the
channel, $P_h$ is dissipated into the lattice through electron-lattice interaction. From here the heat gets removed efficiently by the dilution refrigerator. In this manner we establish a locally enhanced electronic temperature in the channel while the rest of the 2DES remains approximately at base temperature. Using the thermopower of the QPC as a thermometer \cite{Molenkamp1990} we estimate that for the given $I_h$, $T_{el}$ in the channel increases by $\Delta T \approx 30$ mK. We note that because $I_h$ gets modulated with $\omega$, the temperature in the heating channel
oscillates with $2\omega$ since the dissipated power $P_h \propto I_h^2 \propto sin^2 (\omega t) \propto cos(2\omega t)$. This provides all temperature driven effects with a clear signature of an
oscillation frequency of $2\omega$. The thermovoltage $V_{th}$ of the QD is obtained by measuring the potential difference between the contacts of the two cold reservoirs $V_{ref}$ and $V_C$ using a Lock-In amplifier operating at $2\omega = 26 Hz$. Since the QPC is adjusted to a conductance plateau its contribution
to the $V_{th}$ is zero. Hence the measured signal can be attributed fully to the QD. In order to suppress any potential fluctuations at $\omega$ in close vicinity to the QD structure, which may occur due to unwanted capacitive coupling inside the sample, we let the excitation voltage for the
heating current at both contacts of the heating channel oscillate symmetrically with respect to ground. Since reservoir REF is kept grounded, this suppresses oscillations of the electrical potential at
$\omega$ around the QD structure.

\section*{Data Availability} 
The datasets generated and analysed in the study are available upon request from the corresponding authors.


\section*{Acknowledgements}
We thank L. Maier for sample fabrication and D. Reuter and A.D. Wieck (Ruhr-University Bochum) for providing the hetero structure. The work in Würzburg has been funded by the Deutsche Forschungsgemeinschaft DFG (SPP1386). We thank P. Moca and G. Zarand for scientific discussions and help with the DM-NRG code. Y.M. acknowledges support from ISF grant 292/15. HT acknowledges funding through the European research council (Grant No. 339306, METIQUM). AG and YM acknowledge the KITP at the University of California at Santa Barbara, where this project has initiated,  supported in part by NSF Grant No. PHY17-48958, NIH Grant No. R25GM067110, and the Gordon and Betty Moore Foundation Grant No. 2919.01.

\section*{Author contributions statement}
AG, YM and LM initiated the project. AG, YK and YM developed the theory. YK performed the numerical calculations, and the experiments were performed by HT, HB and LM. All authors contributed to the writing of the paper.
\section*{Competing interests}
The authors declare no competing interests.

\end{document}


\flushleft{\huge\textbf{ Supplementary Information}}
\renewcommand{\figurename}{Supplementary Figure}

\flushbottom
\section*{Supplementary Note 1: Derivation of Eqs.~2 and 3}
Eqs.~2 and 3 of the main text express the conductance and T of the full interacting system with arbitrary spectrum and wavefunctions, in terms of these quantities for the non-interacting system, an expression which is valid in weak-coupling($\Gamma\ll T$), Coloumb-blockade  ($U\gg T$) regime. Following Eq.~1 of the main text, we can generally write the expressions for conductance and TR
\begin{align}
\tag{1}
\label{fullform}
G(\mu,T)&=\sum_{ij}G_{ij}=\sum_{ij} \frac{e^2}{h}\mathcal{T}_{ij}^{(0)}\times(P_i^{(N)}+P_j^{(N+1)})\frac{d f(E_j^{(N+1)}-E_i^{(N)}-\mu,T)}{d \mu}\nonumber\\
\mathrm{TR}(\mu,T)&=\sum_{ij}\mathrm{TR}_{ij}=\sum_{ij}\frac{e}{h}\mathcal{T}{ij}^{(0)}\times(P_i^{(N)}+P_j^{(N+1)})\frac{d f(E_j^{(N+1)}-E_i^{(N)}-\mu,T)}{d T}\nonumber
\end{align}
where $\mathcal{T}^{(0)}_{ij}=\sum_{n,m}\Gamma_{ij}\langle\psi_j|d^\dagger_n|\psi_i\rangle\langle\psi_i|d_m|\psi_j\rangle$ and the sum is over all states $i$ and $j$ in the $N$ and $N+1$-electron subspaces, respectively. In the above, we used the fact that for large $U$ the transport properties only involve, at the most, two $N-$electron subspaces, say $N$ and $N+1$. In equilibrium, the probability that the system is in a specific $N$ or $N+1$ many-body state is
\begin{equation}\tag{2}
P_i^{(N)}(\mu,T)=\frac{1}{Z(\mu,T)}\exp({-\frac{E_i^{(N)}-N\mu}{T}}),\quad P_j^{(N+1)}(\mu,T)=\frac{1}{Z(\mu,T)}\exp({-\frac{E_j^{(N+1)}-(N+1)\mu}{T}})
\end{equation}
where $Z=\sum_i\exp\left[(E_i^{(N)}-N\mu)/T\right]+\sum_j\exp\left[(E_j^{(N+1)}-(N+1)\mu)/T\right]$ is the partition function.

\begin{figure}[h]
	\includegraphics[width=0.4\textwidth]{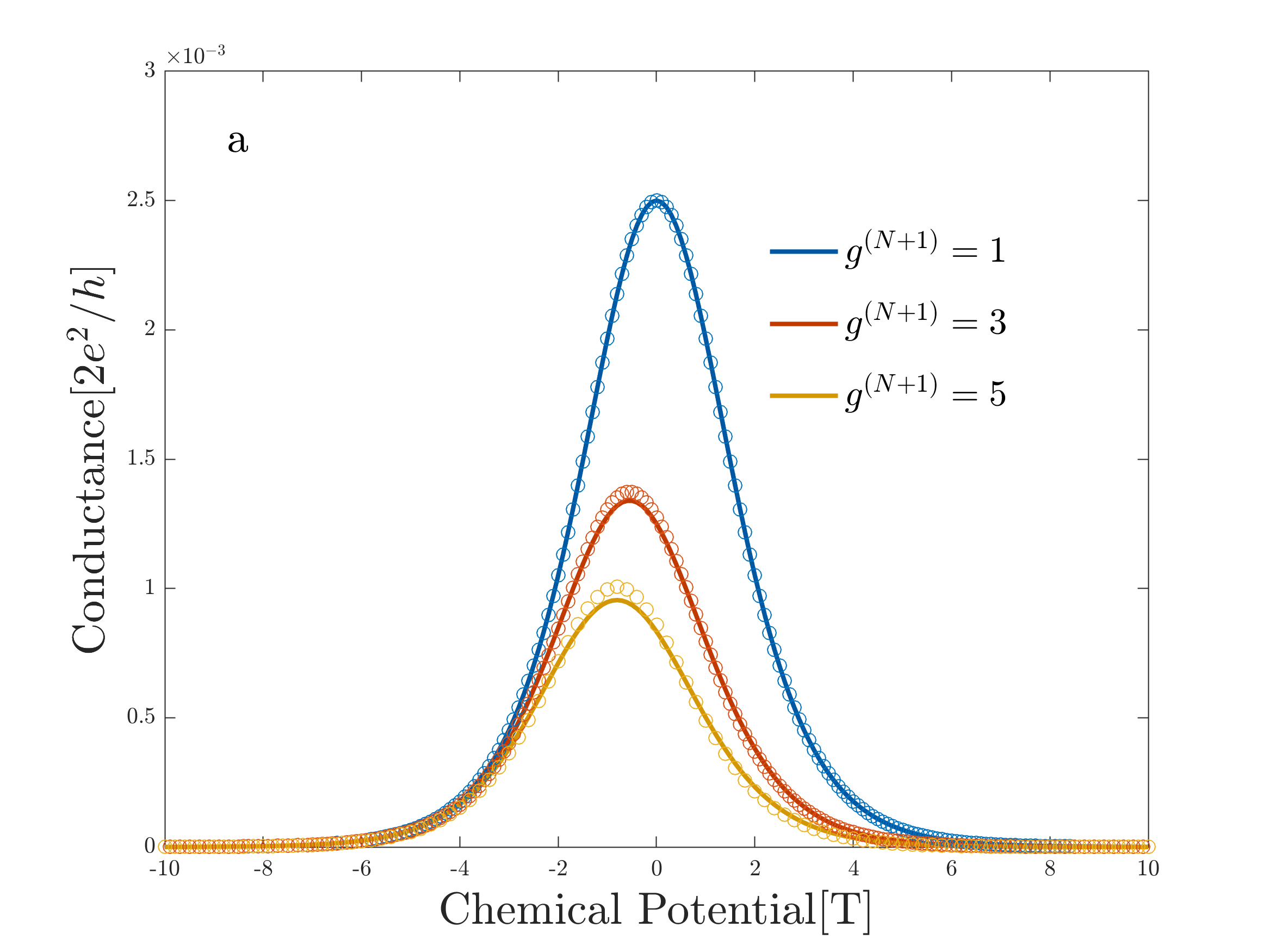}
		\includegraphics[width=0.4\textwidth]{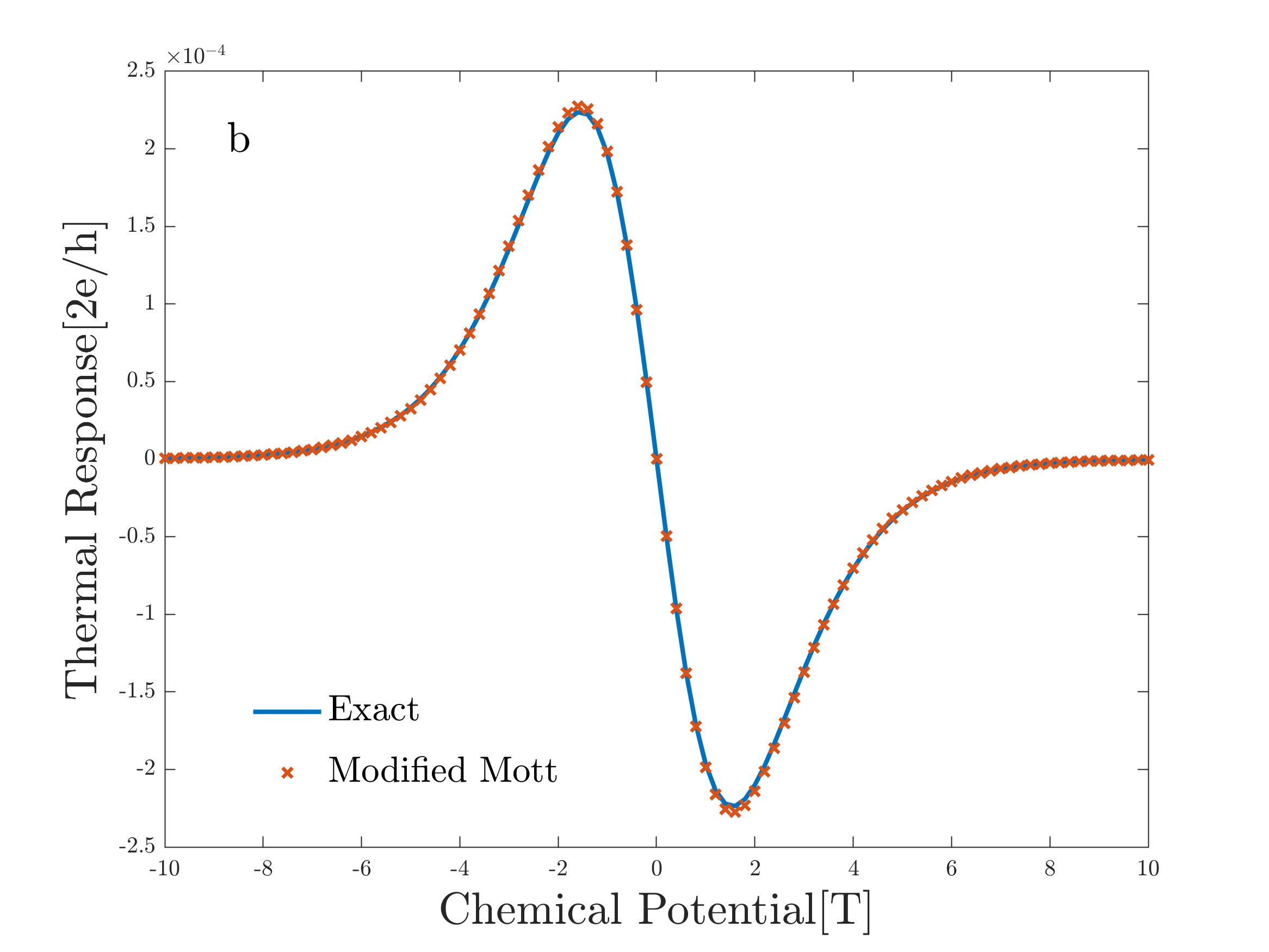}
		\caption{Validity of Supplementary Eq.~\ref{eq:conductance} and the high temperature Mott relation (Supplementary Eq. ~\ref{S7}). (a) Conductance through a quantum dot as a function of chemical potential in the vicinity of the $N\rightarrow N+1$ peak for various degeneracies $g^{(N+1)}$ with $g^{(N)}$ kept at 1. The solid line is the exact calculation (Supplementary Eq.~\ref{fullform}) while the circles are a calculation with an effective Fermi function (Supplementary Eq.~\ref{eq:conductance}). The prefactor C(T) for every curve was found by fitting. (b) The high-temperature Mott relation illustration: direct calculation of TR (solid line) and  calculation of TR using the modified Mott relation (crosses) as a function of chemical potential (in units of temperature) in a system with a single level with energy $\epsilon=0$.}

	\label{fig:eff_fermi}
\end{figure}

We define the function $K_{ij}(\epsilon-\mu,T)$ by the relation
\begin{equation}
\tag{3}
\frac{d}{d\mu}K_{ij}(\epsilon-\mu,T)= (P_i^{(N)}+P_j^{(N+1)})\frac{d f(\epsilon-\mu,T)}{d \mu}.
\end{equation}
The main observation is that, since the factor $(P_i^{(N)}+P_j^{(N+1)})$ approaches a constant in the limits $\mu\rightarrow\pm\infty$, and $f(\epsilon-\mu,T)$ drops, as a function of $\mu$, from unity to zero on a scale of $T$,  the function $K$ will also change from a constant to zero on the scale of $T$, and thus can be well described by a, possibly shifted, Fermi function, $K_{ij}(\epsilon-\mu,T)\simeq C(T) f(\epsilon-\mu-\Delta_{ij}(T),T)$. This Ansatz allows one to write, for the relevant conductance term
\begin{equation}
\tag{4}
G_{ij}(\mu,T)\simeq C(T)\mathcal{T}^{(0)}_{ij}\frac{d}{d\mu}f(E^{(N+1)}_j-E^{(N)}_i-\mu-\Delta_{ij}(T),T).
\label{eq:conductance}
 \end{equation}
Supplementary Eq.~\ref{eq:conductance} is identical, up to a multiplicative constant, to the conductance of a non-interacting system with the same spectrum and matrix elements, but a shifted chemical potential (Eq.~2 of the main text).
In order to find $\Delta_{ij}(T)$ one has to compare the positions of the peaks of $dK/d\mu$ and $df/d\mu$. Before deriving it for the general case, let us look  at a simple case -- an $M$-level QD, in which all the $M$ single-electron levels are degenerate with energy $\epsilon$ and repel one another by the Coulomb repulsion $U$. In this system the energy of the $N$-electron many-body state is $E^{(N)}=N \epsilon+N(N-1)U/2$, and he degeneracy of each many-body energy is $g^{(N)}=\binom{M}{N}$ (the $i,j$ indices have been omitted, since all the degenerate states were assumed to have the same coupling).  The partition function is given by $Z=g^{(N)} \exp\left[(E^{(N)}-N\mu)/T\right]+g^{(N+1)}\exp\left[(E^{(N+1)}-(N+1)\mu)/T\right]$. The equation for the  function $K$ can be solved analytically to give $K(\epsilon,T)=\log\Big(1+\big[h-1\big]f(\epsilon,T)\Big)/\log(h)$, with $h=(g^{(N+1)}/g^{(N)})$. This function can be well approximated by $f(x-\Delta(T),T)$, with $\Delta(T)=T\log(h)/2$. Supplementary Fig. \ref{fig:eff_fermi}a depicts the first Coulomb blockade peak in the conductance through such a QD, with varying degeneracies, using the full expression (Supplementary Eq.~\ref{fullform}) and the Ansatz (Supplementary Eq.~\ref{eq:conductance}) with excellent agreement, though there are noticeable deviations for degeneracy ratios $h\gtrsim5$, which are usually physically irrelevant.

Following the same procedure for the general case, one can again calculate $\Delta_{ij}$ analytically, which results in Eq.~3 of the main text.

While the expression for the conductance looks exactly like that of a non-interacting system with transmission coefficient $\mathcal{T}^{(0)}_{ij}$ and a shifted chemical potential, $\mu\rightarrow \mu+\Delta_{ij}(T)$, the temperature dependence of the shift $\Delta$ will lead to an additional contribution to the TR from $d\Delta_{ij}(T)/dT$ (Eq.~4 of the main text). It is this additional contribution that allows us to determine the entropy difference (see main text and below).
Explicitly, the TR can be treated similarly to the conductance (Supplementary Eq.~\ref{eq:conductance}), where the derivative with respect to $\mu$ is replaced by derivation with respect to temperature. This can be rewritten as

\begin{equation}
\tag{5}
\mathrm{TR}_{ij}(\mu,T)\simeq C(T)\mathcal{T}_{ij}^{(0)}\frac{
\partial}{\partial T}f(E^{(N+1)}_j-E_i^{(N)}-\mu-\Delta_{ij}(T),T)+\frac{d\Delta_{ij}(T)}{dT}C(T)\mathcal{T}_{ij}^{(0)}\frac{d}{d\mu}f(E^{(N+1)}_j-E^{(N)}_i-\mu-\Delta_{ij}(T),T).
\label{eq:TR}
\end{equation}
where the partial derivative in the first term indicates that the shift in the chemical potential is treated as a constant in T. The first term can be identified as a non-interacting TR shifted along the chemical potential and the second term can be identified as the non-interacting conductance times $d\Delta_{ij}(T)/dT$, which results in Eq.~4 in the main text.

\section*{Supplementary Note 2: Derivation of the high-temperature Mott formula}
The standard Mott relation \cite{Cutler1969}  relates the derivative of the conductance to the TR at low temperatures. This relation relies on an approximate relation, valid at low temperatures, between  $df(\epsilon-\mu,T)/d T$, which appears in the expression for the TR, to $d^2f(\epsilon-\mu,T)/d \mu^2$, which, for the non-interacting system, is related to $dG/d\mu$. The relation between these two functions relies on the Sommerfeld expansion, which assumes that the temperature is the smallest energy scale in the problem. In the present case, where $T\gg\Gamma$, this relation has to be modified. In the following we derive an alternative relation between these two functions, valid in this regime, which will then be used to obtain an estimate of the TR for the non-interacting system. As mentioned above, and in the main text, the deviation of the true TR from the non-interacting TR, estimated by the high-temperature Mott formula, allows us to determine $d\Delta/dT$, and as a consequence, the entropy.

One can expand the Fermi function as follows
\begin{align}
\tag{6}
d^2f(\epsilon-\mu,T)/d \mu^2&=\frac{x}{8 T^2}-\frac{x^3}{24T^2}+O(x^5)\nonumber\\
df(\epsilon-\mu,T)/d T&=\frac{x}{4 T}-\frac{x^3}{16T}+O(x^5)\nonumber
\end{align}
where $x=(\epsilon-\mu)/T$.
These two functions can be made equal up to third order in $(\epsilon-\mu)/T$ by determining two parameters in the relation between these two functions such that $df(\epsilon-\mu,T)/d T=T \gamma_1 d^2f(\epsilon-\mu,\gamma_2 T)/d \mu^2$. These factors are an overall factor $\gamma_1=2 \gamma_2^3$ and a temperature factor $\gamma_2=\frac{2}{\sqrt{3}}$. Both these factors do not depend on the system in question, as they only depend on the properties of the Fermi function. As a result of this relation, the high-temperature Mott relation, relying on the fact that transport coefficients are solely governed by their respective Fermi functions and valid only for non-interacting systems, is given by:
\begin{equation}
\tag{7}
\mathrm{TR}^{NI}(\mu,T)=\gamma_1 T \frac{d G^{NI}(\mu,\gamma_2 T)}{d \mu} ,
\label{S7}
\end{equation}
which is quoted as Eq.~5 in the main text.
Supplementary Fig. \ref{fig:eff_fermi}b illustrates the excellent agreement between the real $\mathrm{TR}$, calculated using Supplementary Eq.~\ref{fullform}, and the $\mathrm{TR}$ derived from the calculated conductance through the high temperature Mott relation adaptation, for a noninteracting system.
This relation allows us to estimate the first term on the RHS of Eq.~4 of the main text from measurable quantities, such as the conductance. Note, however, that the relation (\ref{S7}) relates $\mathrm{TR}^{NI}(T)$ to the conductance at temperature $\gamma_2T$.

\section*{Supplementary Note 3: Fitting to numerical and experimental data}
The Mott type contribution (Eq.~5 and Supplementary Eq.~\ref{S7}) relates the TR at temperature $T$ to the derivative of the conductance $G$ at temperature $\gamma_2 T$. Since the actual $G$ and TR, at temperature $T$ and chemical potential $\mu$, are related to the non-interacting $G$ and TR at temperature $T$ and chemical potential $\mu+\Delta(T)$, this leads to a small relative chemical potential shift $\Delta(\gamma_2 T)-\Delta(T)$ between these quantities.  Since $\Delta(T)$ can be deduced from the fitting parameter $A(T)$, this shift can be included self-consistently in the fitting procedure. In addition, the temperature dependent prefactor $C(T)$ in Eq.~2 of the main text, introduces a small factor between the two transport coefficients: $C(\gamma_2 T)/C(T)$. We found that this factor was very close to unity and we did not fit it in our procedure, since it only improves the result negligibly.

Explicitly, the fitting procedure delineated in the main text can be summarized with the fitting formula:
\begin{equation}
\tag{8}
\mathrm{TR}(\mu,T)=\gamma_1\frac{\partial }{\partial T}G(\mu-\big[\gamma_2 T A(\gamma_2 T)-T A(T)\big],\gamma_2 T)+A(T)G(\mu,T)
\label{eq:general}
\end{equation}
where the only fitting parameter, for every single value of temperature, is $A(T)$.

We carried out this procedure separately  in the vicinity of each peak , as the value of $A$ is different for each peak. The "vicinity of each peak" is defined between the points where the TR and conductance vanish.
\begin{figure}
	\includegraphics[width=0.5\textwidth]{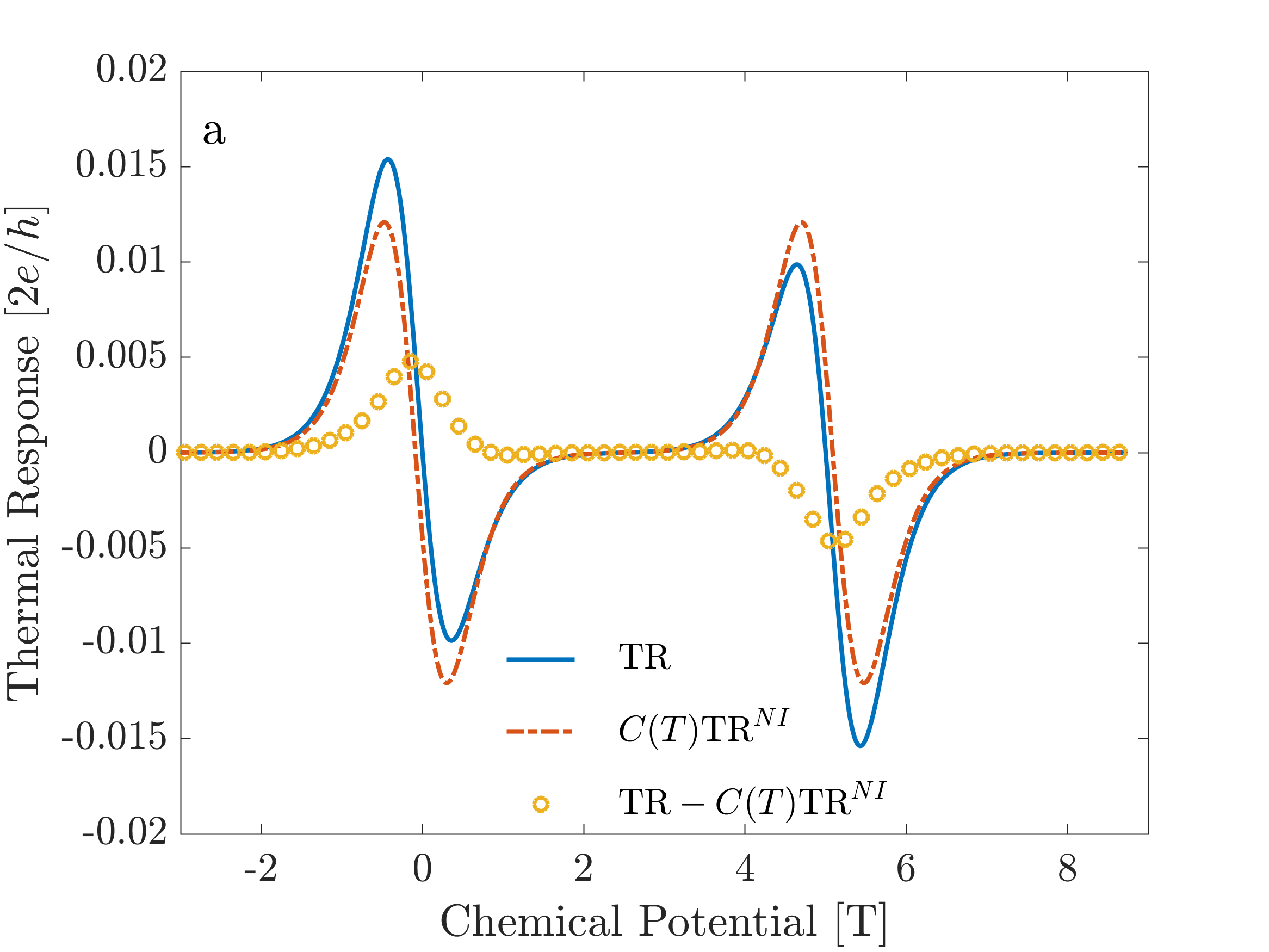}
	\includegraphics[width=0.5\textwidth]{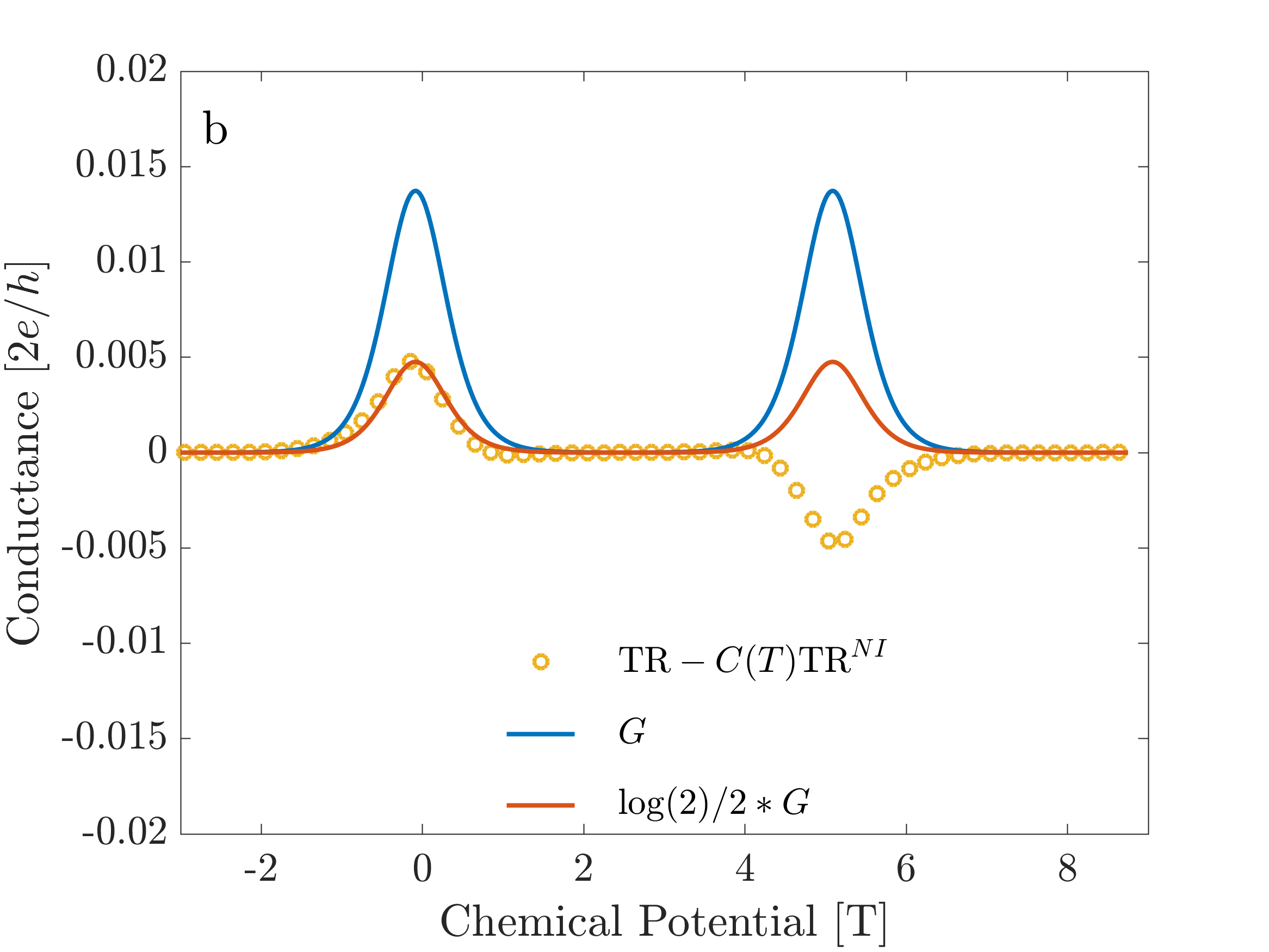}
	\caption{The different components of the thermal response (Supplementary Eq.~\ref{eq:TR}). (a) the full TR (solid blue line), Mott contribution (dot-dash red line) and their difference (yellow circles), illustrating the various terms in Supplementary Eq.~\ref{eq:general}. (b) Comparison of the difference between the full TR and the Mott contribution, and the scaled conductance, as per the second term of the RHS of Supplementary Eq.~\ref{eq:general} }
	\label{fig:parts}
\end{figure}
In order to illustrate the various components of Supplementary Eq.~\ref{eq:general} (and Eq.~4 in the main text), we plot  in Supplementary Fig.~\ref{fig:parts}a  the full TR (solid blue line) for the case of a doubly degenerate level with $U\rightarrow\infty$,  along with the fitting of the first term on the RHS of Supplementary Eq.~\ref{eq:general}, the high temperature Mott contribution (the dot-dash red line). The difference between what would be expected in a non-interacting system (the Mott contribution) and full TR (shown in yellow circles) corresponds to the second term on the RHS of Supplementary Eq.~\ref{eq:general}, which is evidently just a numerical factor times the conductance, as seen in Supplementary Fig.~\ref{fig:parts}b. That numerical factor is $A$, which in this case is half the entropy change across each peak.

In the case of a temperature dependence in $A(T)$, such as that depicted in Figs.~1e,f, the self consistent procedure involves an iterative procedure. We infer $A(T)$ neglecting the contribution of $dA(T)/dT$ to the shift, and then use that inferred value to estimate the derivative needed to calculate the correct shift in chemical potential. In theory, one may continue to iterate, however we find that due to the small effect of this derivative, one iteration is enough, as seen in Figs.~1e,f.

The experimental data and its limitations has offered further challenges. Since experimentally, the two measurements: conductance and thermovoltage, involve different sets of gate voltages, there is an arbitrary shift of the $x$-axis, and we have added this value as a fitting parameter. In addition, the temperature that was experimentally established without current heating is not accurate for the case of an existing temperature bias, so we also added that as a fitting parameter.  The resulting fitting equation reads

\begin{equation}
\tag{9}
\mathrm{TR}(\mu)=\gamma_1\frac{\partial }{\partial T}G(\mu-\beta-(\gamma_2-1) T A)+G(\mu-\beta)A
\label{eq:exp}
\end{equation}
where $A,\beta, T$ are the fitted parameters.

The peaks in the experimental data are not clearly separated. Using different definitions of "peak vicinity" we get slightly different results. In Supplementary Fig.~\ref{fig:TPfit} we depict two extreme choices for peak definition, in (a) we show the fit for the whole peak, up to the mid point between the peaks, and in (b) where only the bulk of the peak is fitted. Such fits allow us to determine the extreme values of ratio of entropy changes, which is independent of the value of $\Delta T$, $\Delta S_1/\Delta S_2=-2.07\pm0.12$. Assuming that this corresponds to the likely scenario of $\Delta S_1=\log 4$ and $/\Delta S_2=-\log2$  allows us to deduce $\Delta T= 20.5mK\pm0.5mK$.

\begin{figure}
	\includegraphics[width=0.5\textwidth]{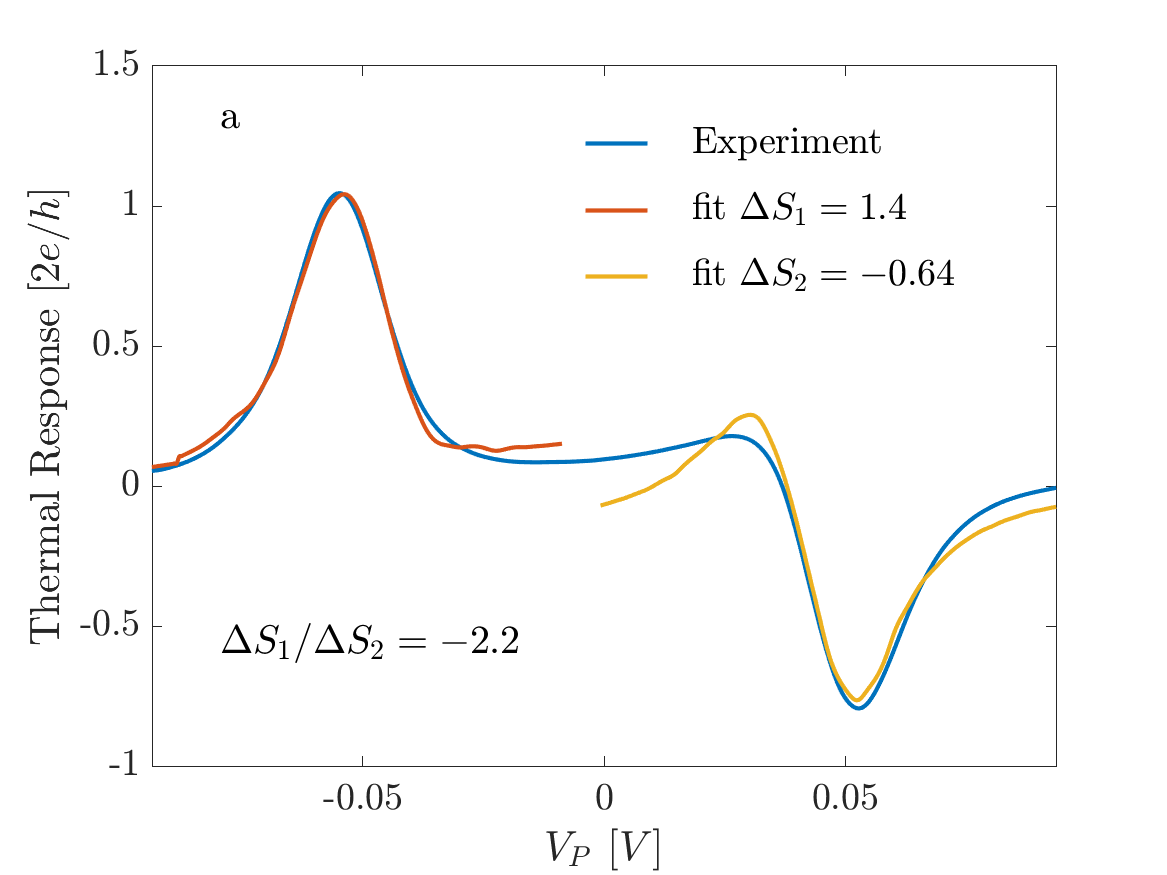}
		\includegraphics[width=0.5\textwidth]{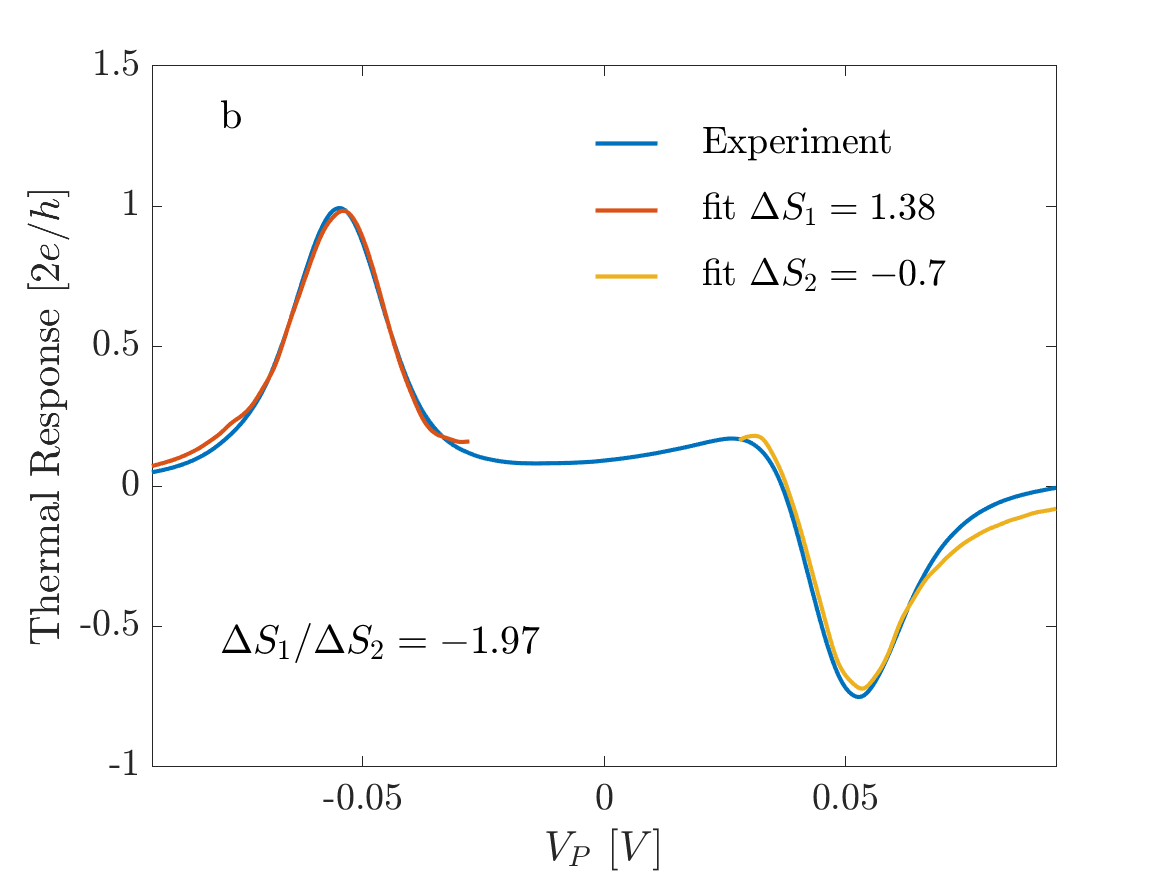}
		\caption{Estimation of variance in fitting procedure. Fitting results for experimental TR similar to Fig.~3c in the main text with two extreme choices for fitting regions around the peaks: (a) Fitting the full range to mid point between the peaks and (b)  fitting only the bulk of the peak.}
		\label{fig:TPfit}
\end{figure}

\section*{Supplementary Note 4: The case with several conducting levels}
Eq.~3 of the main text expresses the shift in the chemical potential for a given transition from state $i$ to state $j$, $\Delta_{ij}$, as a sum of two terms, the first proportional to the energy difference between the two states, while the second, one half of the total free energy, is independent of the particular states $i$ and $j$.
 This shift is obtained, either numerically or experimentally, by dividing the difference between the full TR and that of the corresponding non-interacting system, deduced via the Mott relation, by the conductance. Sec.~3 of the main text considered cases where a transition through single level dominated the conductance.
 In order to demonstrate the case of mutli-level transport, let us consider the model discussed in Sec.~3, a two-level QD, of non-degenerate levels, where each of the transitions between an empty QD to one of the levels being occupied has a finite contribution to conductance, $G_{11}(\mu,T)$ and $G_{12}(\mu,T)$, respectively. In that case, the procedure will approximately yield $\Delta(T) = \big[G_{11}(\mu,T) \Delta_1(T) + G_{12}(\mu,T) \Delta_2(T)\big]/\big[G_{11}(\mu,T)+G_{12}(\mu,T)\big]$, where $\Delta_i(T)$ is the chemical-potential shift corresponding to the transition through level $i$ (note that $\Delta_2-\Delta_1=\Delta\epsilon/2$. Taking the derivative with respect to $T$, we find, in addition to the entropy change term $ \Delta S$ derived before, an additional term
\begin{equation}
\tag{10}
\frac {d\Delta(T)}{dT}  = \Delta S+ \frac{\Delta\epsilon}{T^2}\left(G_{12}(\mu,T)\frac{dG_{11}(\mu,T)}{dT}-G_{11}(\mu,T)\frac{dG_{12}(\mu,T)}{dT}\right)
|_{\mu=\mu_{max}}
\end{equation}
The second term  vanishes when one of the levels dominates the transport, or when the levels are degenerate, or at high temperatures ($T\gg\Delta\epsilon$) or at low temperatures ($T\ll\Delta\epsilon$). So the maximal deviation is expected at $T\simeq\Delta\epsilon$, when the levels have the same coupling to the leads. In this case the deviation is approximately $(\Delta\epsilon/\Delta T)^2/(1+\cosh(\Delta\epsilon/T))$.

\begin{figure}[h]
	\includegraphics[width=0.5\textwidth]{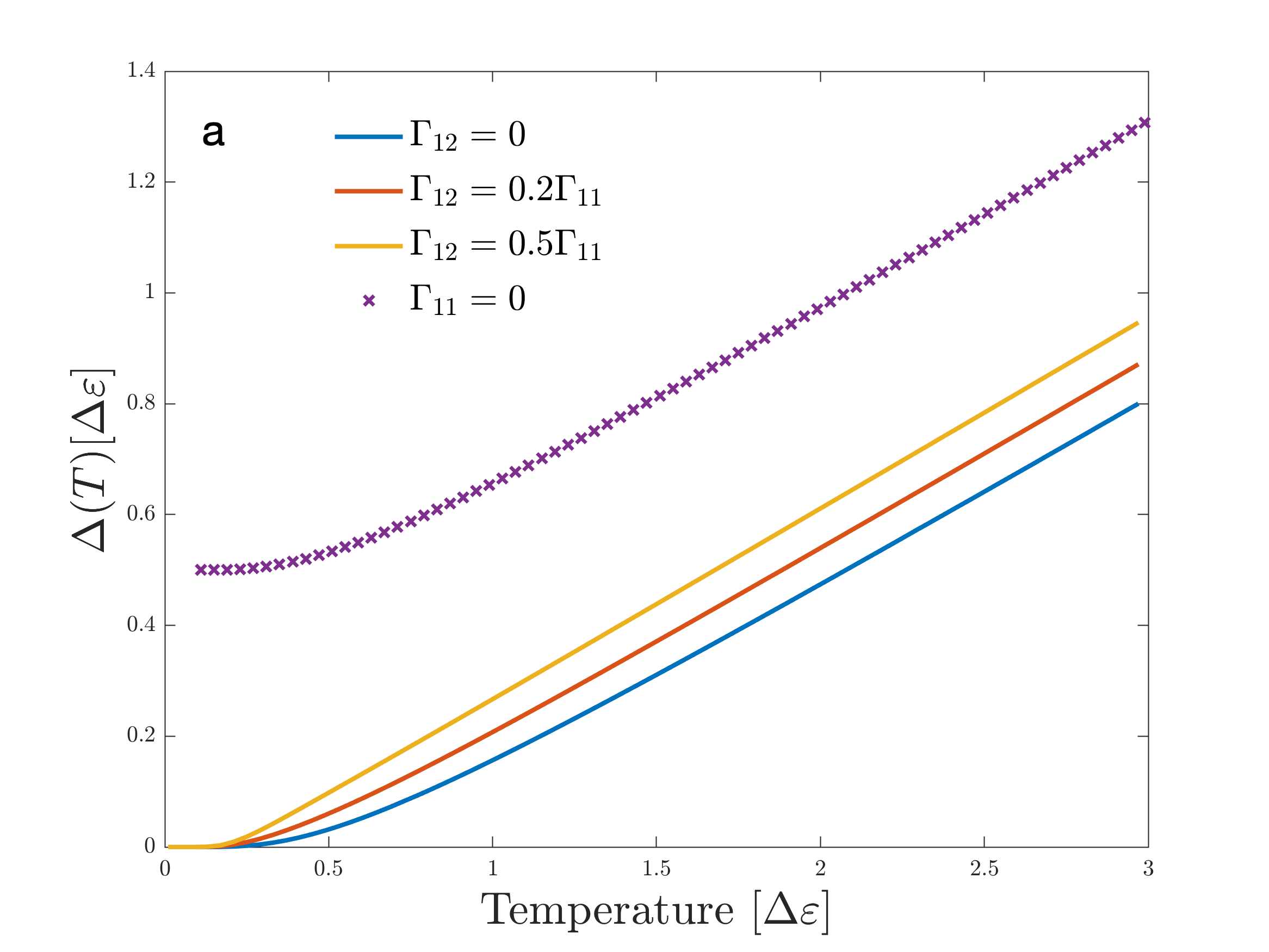}
	\includegraphics[width=0.5\textwidth]{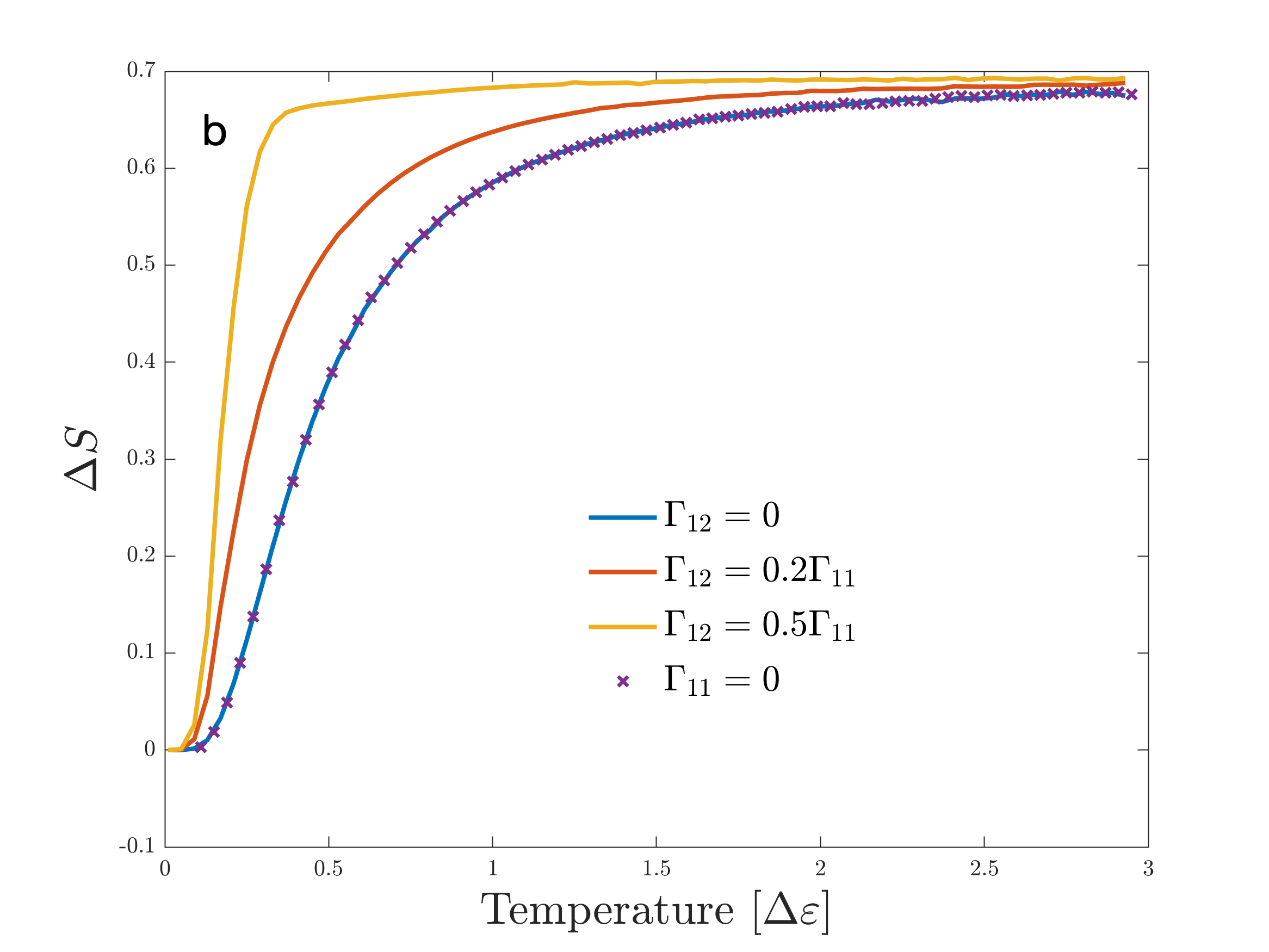}
	\caption{Multi-level fitting results. Fitting results for (a) $\Delta(T)$ and (b) the deduced entropy change $\Delta S$ for various ratio of coupling $\Gamma_{ij}$ to the leads. The correct result for deduced entropy is recovered in the two cases of single dominating transition.}
	
	\label{fig:nondeg}
\end{figure}

Supplementary Fig.~\ref{fig:nondeg}a depicts the deduced $\Delta(T)$ for the two-level case, with different ratios of couplings to the leads. As expected, for intermediate values of the ratio of the couplings, one finds that the high-$T$ value of $\Delta S$ is a weighted average of the $\Delta S_{11}$ and $S_{12}$, while at low temperatures, they all converge to $S_{11}$. Clearly, in the regime $T\simeq\Delta\epsilon$ the derivative becomes larger. This is manifested in Supplementary Fig.~\ref{fig:nondeg}b, where the deduced values of the entropy change are depicted, where the exact result for entropy change is captured for the two cases with only one contributing transition ($\Gamma_{12}=0$ or $\Gamma_{11}=0$). When both level contributes, some deviations from the exact results are noticeable. Note, however, that, as the couplings to the leads depend exponentially on energy, it is unlikely that they will be equal to each other, and thus this regime may not be physically relevant.

\section*{Supplementary Note 5: Experimental setup and procedure}
The device used in the experiments was fabricated from a GaAs/AlGaAs heterostructure hosting a two dimensional electron system (2DES) 90 nm under the surface. The 2DES has a nominal carrier density $n = -2.98\times 10^{11} cm^{-2}$ and a mobility $\mu = 550 \times 10^5 cm^2(Vs)^{-1}$ at T=4 K.
The gate pattern was fabricated with standard optical and e-beam lithography and lift-off techniques. The gate electrodes consist of a 2 nm thick Ti adhesion layer and a 50 nm thick Au layer. Ohmic contacts with the 2DES were obtained through annealed Au/Ge pads and bonding.

\begin{figure}
	\includegraphics[width=1\textwidth]{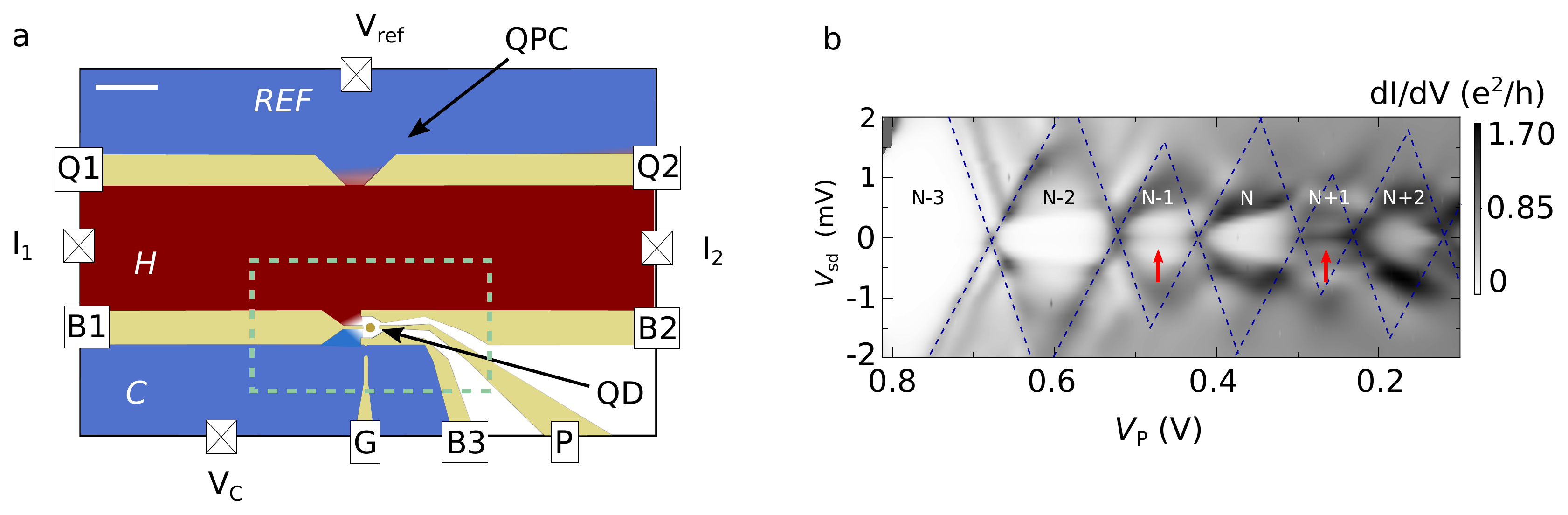}
	\caption{Experimental setup and analysis. (a) Gate layout of the sample. The gates (yellow) B1, B2, B3 and P are used to confine the QD (yellow dot), which is tunnel coupled to the equilibrium electron reservoirs H (red) and C (blue). Reservoir H is shaped through the pairs of gates B1, B2 and Q1, Q2 into a channel. Gates Q1 and Q2 further form a quantum point contact used for thermometry. For thermopower measurements a heating current $I_h$ is applied to the heating channel through contacts $I_1$ and $I_2$. The thermovoltage of the QD is measured using the voltage probes $V_{ref}$ and $V_C$, which are both 	connected to cold reservoirs. The scale bar corresponds to 1 $\mu$m. The dashed frame denotes the region of which a SEM image is shown in the main text. (b) Differential conductance dI/dV stability 	diagram of the QD with all reservoirs at low temperature. The Coulomb diamonds are indicated with 	dashed lines. The charge occupation number associated with the respective diamonds are denoted	(N-3), (N-2), …etc. Red arrows denote regions with Kondo enhanced zero bias conductance,	indicative for an odd electron occupation number.
		}
\label{fig:Sexp}
\end{figure}

Supplementary Fig.~\ref{fig:Sexp}a depicts a drawing of the gate pattern of the sample. The scale bar corresponds to 1 $\mu m$. Gates are shown in yellow. Reservoirs which are at low ($T_c$) and higher ($T_h = T_c + \Delta T$) temperatures during thermopower measurements are denoted with blue and red color, respectively. The green dashed square indicates the region of which an SEM image is shown in the main text. The quantum dot is formed with gates B1, B2, B3, and P. The heating channel H (red) (width $w = 20 \mu m$ and a length of $20 \mu m$) serves as one of the leads of the QD. It is equipped with two Ohmic contacts $I_1$ and $I_2$ through which a heating current $I_h$ is applied to the channel. The other lead of the QD, labeled C, is equipped with a voltage probe $V_C$. It serves as a cold equilibrium reservoir (blue). The quantum point contact (QPC) formed by the gates Q1 and Q2 is placed exactly opposite to the QD. The QPC is adjusted to the 10 $e^2/h$ conductance plateau. It separates the heating channel from the cold electron reservoir denoted REF. The thermovoltage $V_{th}$ measured between $V_{ref}$ and $V_C$ as a response to a temperature increase in H is then given by $V_{th} = V_C – V_{ref} = (S_{QD} – S_{QPC})\Delta T$. Since the QPC is adjusted to a conductance plateau, its thermopower $S_{QPC} = 0$. Therefore the measured voltage $V_{th}$ can be assigned entirely to the thermopower of the QD, $S_{QD}$.

For a heating current $I_h = 70$ nA the electron gas in the channel typically heats up by approximately $\Delta T \approx 50 mK$ (cf. supplementary section C in Refs. \cite{Thierschmann2015,thierschmann2014heat}). If we take into account that the electron density in our material is higher by approximately a factor 1.4 compared with Ref. \cite{Thierschmann2015} ($2.98 \times 10^{11} cm^{-2}$ here and $2.14 \times 10^{11} cm^{-2}$ in Reference \cite{Thierschmann2015}) and that the heat capacity of a 2DES follows the carrier density linearly in first approximation \cite{Molenkamp1990}, we estimate that for $I_h = 70 nA$, $\Delta T \approx 30 mK$ in our sample.

Supplementary Fig.~\ref{fig:Sexp}b shows the stability diagram obtained from measurements of the differential conductance dI/dV of the QD with all reservoirs at low temperature. The Coulomb diamonds, signatures of a fixed charge occupation number N of the QD, are highlighted with dashed blue lines. From the size of the Coulomb diamonds on the $V_{sd}$ axis we can extract the charging energy $U$ of the QD, $U \approx 1.7 meV$. This allows us to calculate the electrostatic lever arm of the plunger gate voltage $\alpha = 0.016 e$ which can then be used to convert the plunger gate voltage axis $V_P$ into QD energies $\epsilon $ using $\epsilon = \alpha \times V_P$. From Fig. S3b we see that in the Coulomb blockade regions for (N-1) and (N+1) the conductance for small bias voltage is not fully suppressed but it shows a zero bias conductance. This is a signature of Kondo correlations being present for these charge configuration.
\section*{Supplementary Note 6: Role of magnetic field }

The experimental conductance and thermovoltage data discussed in the main text were measured with a perpendicular magnetic field B = 0.6 T applied. In Supplementary Fig.~\ref{fig:SB} we show the evolution of the conductance G with magnetic field as a function of plunger gate voltage $V_P$. We see that for B  = 0 (yellow line), the conductance G  is close to zero for higher occupation numbers ($V_P > 0.1 V$). For smaller occupation numbers ($V_P < 0.1 V$), corresponding to the part discussed in the main text, two conductance peaks are visible. As B is increased, transmission through the QD changes dramatically. For $V_P > 0.1 V$, two conductance peaks emerge, with a partly suppressed Coulomb blockade valley in between. For $V_P < 0.1 V$ the effect of B on QD transmission is less drastic, yet clearly visible from changes in peak shape and height. Similar behavior has been observed for the B dependent transmission of a QD, for example, by van der Wiel et al. [21]. At B = 0.6 T (blue line), G has evolved into a clear, well-defined series of conductance peaks. In order to base our thermopower analysis on solid footing, and remove contributions from QD states with unusually suppressed transmission, we have therefore chosen B = 0.6T as experimental condition to study the thermopower of the system.
\begin{figure}
	\includegraphics[width=0.5\textwidth]{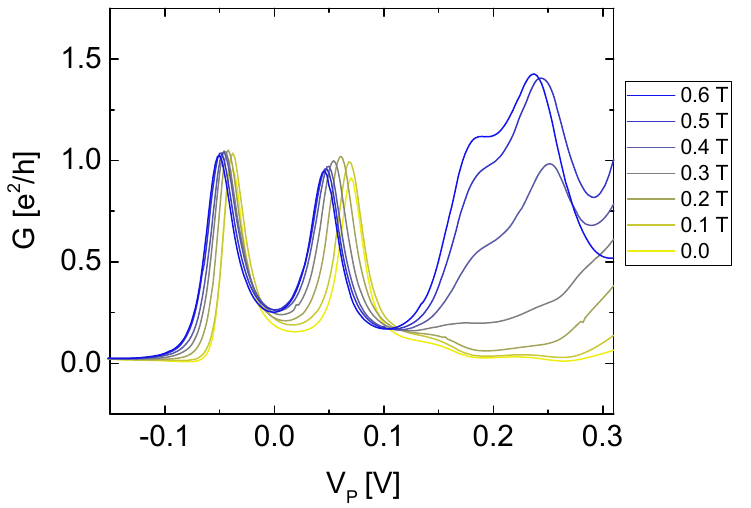}
	\caption{Magnetic field dependence of the data.
		Conductance G as a function of plunger gate $V_P$ for different perpendicular magnetic fields B = 0 to 0.6T. $V_P = 0$ has been set to the center of the conductance valley between the first two conductance peaks for B=0.6T.
	}
	\label{fig:SB}
\end{figure}
\thispagestyle{empty}

\section*{Supplementary References}
\renewcommand{\section}[2]{}%